\documentclass[12pt]{article}
\usepackage{epsfig}
\usepackage{psfrag}
\usepackage{latexsym}
\usepackage{indentfirst}
\usepackage{fancyhdr}
\usepackage{amssymb}
\usepackage{amsmath}
\usepackage{amsfonts}
\usepackage{cite}
\usepackage{bbold}
\usepackage{color}
\usepackage[footnotesize]{caption2}
\usepackage{graphicx}
\usepackage[center,footnotesize,hang]{subfigure}
\usepackage{url}

\makeatletter

\newcommand{\Rmnum}[1]{\expandafter\@slowromancap\romannumeral #1@}
\makeatother

\newcommand{\pa}{\partial}

\newcommand{\td}{\tilde}

\def\be{\begin{equation}}
\def\ee{\end{equation}}
\def\bc{\begin{center}}
\def\ec{\end{center}}
\def\bea{\begin{eqnarray}}
\def\eea{\end{eqnarray}}

\catcode`@=11
\def\marginnote#1{}
\newcount\hour
\newcount\minute
\newtoks\amorpm
\hour=\time\divide\hour by60
\minute=\time{\multiply\hour by60 \global\advance\minute by-\hour}
\edef\standardtime{{\ifnum\hour<12 \global\amorpm={am}%
        \else\global\amorpm={pm}\advance\hour by-12 \fi
        \ifnum\hour=0 \hour=12 \fi
        \number\hour:\ifnum\minute<10 0\fi\number\minute\the\amorpm}}
\edef\militarytime{\number\hour:\ifnum\minute<10 0\fi\number\minute}
\def\draftlabel#1{{\@bsphack\if@filesw {\let\thepage\relax
   \xdef\@gtempa{\write\@auxout{\string
      \newlabel{#1}{{\@currentlabel}{\thepage}}}}}\@gtempa
   \if@nobreak \ifvmode\nobreak\fi\fi\fi\@esphack}
        \gdef\@eqnlabel{#1}}
\def\@eqnlabel{}
\def\@vacuum{}
\def\draftmarginnote#1{\marginpar{\raggedright\scriptsize\tt#1}}
\def\draft{\oddsidemargin 0.0truein
        \def\@oddfoot{\sl preliminary draft \hfil
        \rm\thepage\hfil\sl\today\quad\militarytime}
        \let\@evenfoot\@oddfoot \overfullrule 3pt
        \let\label=\draftlabel
        \let\marginnote=\draftmarginnote
   \def\@eqnnum{(\theequation)\rlap{\kern\marginparsep\tt\@eqnlabel}%
\global\let\@eqnlabel\@vacuum}  }
\catcode`@=12
\begin{document}
\begin{titlepage}
\vspace*{-1cm}
\phantom{hep-ph/***}

\hfill{USTC-ICTS-11-14}

\vskip 2.5cm
\begin{center}
{\Large\bf One Electron Atom in Special Relativity with de Sitter Space-Time Symmetry}
\end{center}
\vskip 0.2  cm
\vskip 0.5  cm
\begin{center}
{\large Mu-Lin Yan}~\footnote{Email: mlyan@ustc.edu.cn}
\\
\vskip .3cm {\it Interdisciplinary Center for Theoretical Study,}
\\
 {\it Department of Modern Physics,}
\\
{\it University of Science and Technology of China, Hefei, Anhui
230026, China}
\end{center}
\vskip 0.7cm
\begin{abstract}
\noindent

\noindent The de Sitter invariant Special Relativity (dS-SR) is a SR with constant curvature, and a natural extension of usual Einstein SR (E-SR). In this paper, we solved the dS-SR Dirac equation of Hydrogen  by means of the adiabatic approach and the quasi-stationary perturbation calculations of QM. Hydrogen atoms are located on the light cone of the Universe.  FRW metric and $\Lambda$CDM cosmological  model are used to discuss this
issue. To the atom, effects of de Sitter space-time geometry
described by Beltrami metric are  taken into account. The
dS-SR Dirac equation turns out to be
 a time dependent quantum Hamiltonian system. We revealed that: 1,The fundamental
physics constants $m_e,\;\hbar,\;e$ variate adiabatically along with cosmologic time
in dS-SR QM framework. But
the fine-structure constant $\alpha\equiv e^2/(\hbar c)$ keeps to be invariant; 2,$(2s^{1/2}-2p^{1/2})$-splitting due to dS-SR QM effects:  By means of perturbation theory, that splitting $\Delta E(z)$ were calculated analytically, which belongs to $\mathcal{O}(1/R^2)$-physics of dS-SR QM. Numerically, we found that when $|R|\simeq \{10^3 Gly,\;10^4 Gly,\;10^5 Gly\;\}$, and $z\simeq \{1,\;{\rm or}\;2\}$, the $\Delta E(z)>> 1{\rm (Lamb\; shift)}$. This indicate that for these cases the hyperfine structure effects due to QED could be ignored, and the dS-SR fine structure effects are dominant. This effect could be used to determine the universal constant $R$ in dS-SR, and be thought as a new physics beyond E-SR.

\end{abstract}
\vskip0.2in


\end{titlepage}
\setcounter{footnote}{1}
 \vskip2truecm
\section{Introduction}
\noindent  Einstein's Special Relativity (E-SR) has global Poincar\'{e}-Minkowski space-time
symmetry.
E-SR indicates the space-time metric is
$\eta_{\mu\nu}=diag\{+,-,-,-\}$. The most general transformation to
preserve metric $\eta_{\mu\nu}$ is Poincar\'e group. It is well
known that the Poincar\'e group is the limit of the de Sitter group
with the sphere radius $R\rightarrow \infty$. Thus people could pursue
whether there exists another type of de Sitter transformation with
$R \rightarrow finite$ which also leads to a Special Relativity
theory (SR).  In 1970's, Lu, Zou
and Guo suggested the Special Relativity theory with de Sitter space-time symmetry (dS-SR) \cite{look}\cite{Lu74}. In recent years, there are various studies of this theory \cite{Guo}. In 2005, Yan, Xiao, Huang and Li performed Lagrangian-Hamiltonian formulism for dS-SR dynamics with two
universal constants $c$ and $R$, and suggested the quantum mechanics of dS-SR \cite{Yan1}.  There is one universal parameter
$c$ (speed of light) in the Einstein's Special Relativity
(E-SR). By contrast, there are two universal parameters
in the de Sitter Special Relativity (dS-SR):
 $c$ and $R$ (the radius of de Sitter
sphere and to character the cosmic radius). In this present paper, we try to study one-electron atoms, typically Hydrogen atom, of a distant galaxy (e.g., a Quasi-Stellar Object (QSO)) by means of dS-SR Quantum Mechanics (QM) suggested in Ref.\cite{Yan1}.

As is well known that one of GR principles is existence of Locally Inertial System (LIS) at any point with small enough vicinity region in the curved space-time. In LIS, the expressions of physics laws are the same as ones in SR. Therefore determining the energy level shifts of a distant Hydrogen atom due to dS-SR QM will be useful to the cosmology when the curved space-time is the Friedmann-Robertson-Walker (FRW) Universe.

Ref.\cite{Yan1} shows that the dS-SR dynamical action for free particle associates the dynamics with time- and coordinates-dependent Hamiltonian. In other hand, the Noether theorem assures the symmetry's Neother charges to be conserved even though that the Hamiltonian is time- and space-dependent. In Ref.\cite{Yan1}, 10 external conserved Neother charges for dS-SR have been explicitly presented, which are free particle's energy, 3 momenta, 3 angular-momenta and 3 boost generators (see, Eqs (52)-(56) in \cite{Yan1}). Thus,   the energy conservation law  in dS-SR holds, and at the same time the dS-SR dynamics is a time-dependent  Hamiltonian system. Contrasting with E-SR dynamics, this is a remarkable feature of dS-SR. This will cause time-dependent level shifts in atomic physics, and lead to some remarkable observable effects in cosmology, for instance, the physics constants varying adiabatically \cite{Yan2} and some specific level shifts for Hydrogen atom caused by time interval bing on cosmic scale. In this paper, we will focus on the splitting effect between $2s^{1/2}$ and $2p^{1/2}$ states of Hydrogen in dS-SR Quantum Mechanics (QM).

In this paper, the adiabatic approach \cite{Born}\cite{Messian}\cite{Bayfield}\cite{KSZ} will be used to deal with the time-dependent Hamiltonian problems in dS-SR QM.
Generally, to a $H(x,t)$, we may express it as
$H(x,t)=H_0(x)+H'(x,t)$. Suppose two eigenstates $|s\rangle$ and $|m
\rangle$ of $H_0(x)$ do not generate, i.e., $\Delta E\equiv \hbar (
\omega_{m}-\omega_s)\equiv\hbar \omega_{ms}\neq 0$.
 The validness of for  adiabatic
approximation relies on the fact that the variation of the potential
$H'(x,t)$ in the the Bohr time-period $(\Delta
T_{ms}^{(Bohr)})\dot{H}'_{ms}=(2\pi/\omega_{ms})\dot{H}'_{ms}$
is much less than $\hbar \omega_{ms}$, where $H'_{ms}\equiv \langle m|H'(x,t)|s\rangle$. That  makes the quantum
transition from  state $|s\rangle$ to  state $|m \rangle$ almost
impossible. Thus,  the non-adiabatic effect corrections are small
enough (or tiny) , and  the adiabatic approximations are legitimate
. To the wave equation of dS-SR QM of atoms discussed in this paper,
we show that the perturbation Hamiltonian described the time evolutions of the system $H'(x,t)\propto (c^2t^2/ R^2)$ (where $t$ is the cosmic time). Since $R$ is
cosmologically large and $R>>ct$, the factor
$(c^2t^2/ R^2)$ will make the
time-evolution of the system is so slow that the adiabatic
approximation works. We shall  provide a calculations to
 confirm this point in the paper.
By means of this approach, we solve the
stationary dS-SR Dirac equation for one electron atom, and
the spectra of the corresponding Hamiltonian with time-parameter are
obtained. Consequently, we find out that the electron mass $m_e$, the electric charge $e$ and the Planck constant $\hbar$ vary  as cosmic time going by, but the fine structure constant
$\alpha=e^2/(\hbar c)$ keeps to be invariant.
 Those are interesting  consequences since they indicate that the time-variations of fundamental physics
constants are due to well known quantum evolutions of time-dependent
quantum mechanics that has been widely discussed for a long history (e.g., see
\cite{Bayfield} and the references within).

The life time of a stable atom, e.g., the Hydrogen atom,
is almost infinitely long. We can practically compare the spectra of
atoms at nowadays laboratories  to ones emitted (or absorbed) from
the atoms of a distant galaxy, typically a Quasi-Stellar Object (QSO). The time interval could be on the
cosmic scales. Such observation of spectra of distant astrophysical
objects may encode some cosmologic information in the atomic energy
levels at the position and time of emission. As is well known that the solutions of E-SR Dirac equation of atom are cosmologic effects free because  the Hamiltonian of E-SR is time-independent, and the solutions at any time
are of the same. Thus, after deducting Hubble red shifts, any deviation of cosmology atom spectrum observations from the  results of E-SR Dirac equation could attribute to some new physics beyond E-SR. dS-SR is one of the most straightforward answers to such kind of deviations.

In E-SR Dirac equation of Hydrogen, $2s^{1/2}$- and $2p^{1/2}$-states are completely degenerate. The hyperfine effects of QED break this degeneracy, and turn out famous ``Lamb Shift": i.e., $E_{2s^{1/2}}-E_{2p^{1/2}}\equiv 1\; Lamb\; Shift=4.35152\times 10^{-6}eV=1057.9MHz\times 2\pi\hbar$. In this paper, we solve the dS-SR Dirac equation for one-electron atom and reveal a dS-SR effect which also contribute a level shift to break the $2s^{1/2}-2p^{1/2}$-degeneracy. This shift is proportional to $Q^1Q^0/R^2$, where $Q^1$ is the distance  between the Earth and an observed galaxy (e.g., a QSO), $Q^0$ is the corresponding time interval, and $R$ is the radius of de Sitter sphere. When $Q^1Q^0/R^2$ were large enough, the $2s^{1/2}-2p^{1/2}$ splitting due to dS-SR effects would be much larger than QED's Lamb shift, the observation of this splitting in cosmological experiments could provide a criteria to check dS-SR.

The contents of the paper are organized as follows: In section II,
we briefly recall the classical mechanics of de Sitter special relativity and the corresponding quantum mechanics. The dS-SR Dirac equation for spin-1/2 is presented;
Section III is devoted to discuss Hydrogen atom described by dS-QM and embedden in light cone of Friedmann-Robertson-Walker (FRW) Universe; Section IV shows the solutions of usual E-SR Dirac equation for Hydrogen atom located at distant galaxy. Especially, the wave functions and energy values of the states $2s^{1/2}$ and $2p^{1/2}$ are presented explicitly. The Hydrogen atom energy level shifts due to gravity in FRW Universe are estimated; In Section V, we derive dS-SR Dirac equation for Hydrogen atom up to terms being proportional to $1/R^2$;
Section VI: dS-SR Dirac equation for spectra of Hydrogen atom; Section VII: Adiabatic approximation solution to dS-SR Dirac spectra equation and time variation of physical constants; Section VIII: $2s^{1/2}-2p^{1/2}$ splitting in the dS-SR Dirac equation of Hydrogen; Finally, in Section IX, we briefly discuss and summarize the results of this paper.  In Appendix
A, we derive the electric Coulomb Law in QSO-Light-Cone Space; In
Appendix B, we show the calculations of adiabatic approximative wave
functions in dS-SR-Dirac equation of Hydrogen in
detail. In Appendix C, we provide analytic calculations to the matrix elements of perturbation Hamiltonian in dS-SR, which yield $(2S^{1/2}-2p^{1/2})$-hyperfine splitting discussed in the text.

\section{Special Relativity with de Sitter Symmetry and dS-SR Dirac equation }

The precise dS-SR theory were formulated in 1970--1974 by LU, ZOU and GUO \cite{look}\cite{Lu74} (for the English version, see, e.g., Ref.\cite{Yan1}\cite{Guo}). Two theorems  were proved:

Lemma I: Inertial motion law for free particles holds to be true in
the space-time characterized by Beltrami metric
\begin{eqnarray}\label{star28}
B_{\mu\nu}(x)={\eta_{\mu\nu} \over \sigma (x)}+{1 \over
R^2 \sigma(x)^2} \eta_{\mu\lambda}\eta_{\nu\rho} x^\lambda x^\rho,
\end{eqnarray}
where $\sigma(x)\equiv 1-{1\over R^2} \eta_{\mu\nu}x^\mu
x^\nu,$  and the constant
$R$ is the radius of the pseudo-sphere in de Sitter (dS) space.
$R^2>0$ or $<0$ that corresponds to dS
symmetries $SO(4,1)$ or $SO(3,2)$ respectively.  This claim means
that in the space-time characterized by $B_{\mu\nu}$, the velocity of free
particle is constant, i.e.,
\begin{equation}\label{motion}
\dot{\bf{x}}=\overrightarrow{v}=constant,~~~~{\rm{for}\;\rm{free\;particle}}
\end{equation}
which is exactly the counterpart  of E-SR's inertial law in
Minkowski space characterized by $\eta_{\mu\nu}$. (see  Refs.
\cite{Guo} \cite{Yan1} for the English version of proof to
Eq.(\ref{motion})).

 Lemma II: The  de Sitter space-time transformation
 preserving $B_{\mu\nu}(x)$ is as follows
\begin{eqnarray}\label{transformation}
x^{\mu}\longrightarrow \tilde{x}^{\mu} &=& \pm \sigma(a)^{1/2}
\sigma(a,x)^{-1}
(x^{\nu}-a^{\nu})D_{\nu}^{\mu}, \\
    \nonumber D_{\nu}^{\mu} &=& L_{\nu}^{\mu}+ R^{-2} \eta_{\nu
\rho}a^{\rho} a^{\lambda} (\sigma
(a) +\sigma^{1/2}(a))^{-1} L_{\lambda}^{\mu} ,\\
\nonumber L : &=& (L_{\nu}^{\mu})\in SO(1,3), \\
\nonumber \sigma(x)= 1&-&{1\over R^2}{\eta_{\mu \nu}x^{\mu}
x^{\nu}},~~  \sigma(a,x)= 1-{1 \over R^2}{\eta_{\mu
\nu}a^{\mu} x^{\nu}}.
\end{eqnarray}
where $ x^{\mu}$ is the coordinate in an initial Beltrami frame, and
$ \tilde{x}^{\mu}$ is in another Beltrami frame whose origin is
$a^{\mu}$ in the original one. There are 10 parameters in the
transformations between them. Under the transformation
(\ref{transformation}), we have the equation  preserving
$B_{\mu\nu}$ as follows
\begin{equation} \label{B01}
 B_{\mu\nu}(x)\longrightarrow \widetilde{B}_{\mu\nu}(\widetilde{x})={\pa x^\lambda \over \pa
 \widetilde{x}^\mu}{\pa x^\rho \over \pa
 \widetilde{x}^\nu}B_{\lambda\rho}(x)=B_{\mu\nu}(\widetilde{x}).
\end{equation}
( see  Appendix of Ref. \cite{Yan1} for the English version of proof
to Eq.(\ref{B01})). Eq.(\ref{B01}) will yield conservation laws for
the  energy, momenta, angular momenta and boost chargers of
particles in dS-SR mechanics \cite{Yan1}. Here, we like to address that the space-time symmetry described by Eqs. (\ref{transformation}) (\ref{B01}) is a global symmetry since both $a^\mu$ and $L^\mu_\nu$ are constants instead of functions of space-time $x^\mu$. This situation is same as E-SR's, where the Poincar\'{e} symmetry is global.

Based on the dS-SR space time theory described in above two lemmas,
the dS-SR dynamics described by the Lagrangian as follows \cite{Yan1}
\begin{equation}\label{27}
 L=-m_0c{ds\over
 dt}=-m_0c{\sqrt{B_{\mu\nu}(x)dx^\mu dx^\nu}\over dt}=-m_0c{\sqrt{B_{\mu\nu}(x)\dot{x}^\mu \dot{x}^\nu}},
 \end{equation}
where $\dot{x}^\mu=\frac{d}{dt}x^\mu$, $B_{\mu\nu}(x)$ is Beltrami
metric (\ref{star28}). Then the canonic momenta and canonic energy (i.e., Hamiltonian) reads
\begin{eqnarray}\label{40}
   \pi_{i} &=&\frac{\pa L}{\pa
\dot{x}^i} = -m_0 \sigma(x) \Gamma B_{i \mu}\dot{x}^{\mu} \\
 \label{40a}    H
&=&\sum_{i=1}^3 \frac{\pa L}{\pa \dot{x}^i} \dot{x}^i
-L=m_0 c \sigma(x) \Gamma B_{0 \mu}\dot{x}^{\mu},
  \end{eqnarray}
where {\begin{eqnarray} \label{new parameter}
 \Gamma^{-1}\hskip-0.1in =\sigma(x) \frac{ds}{c dt}={1\over R} \sqrt{(R^2-\eta_{ij}x^i
x^j)(1+\frac{\eta_{ij}\dot{x}^i \dot{x}^j}{c^2})+2t \eta_{ij}x^i
\dot{x}^j -\eta_{ij}\dot{x}^i \dot{x}^j t^2+\frac{(\eta_{ij}
x^i\dot{x}^j)^2}{c^2}}.
\end{eqnarray}}
From the equation of motion $\delta L=0$, we have \cite{Yan1}
\begin{equation}\label{38a} \dot{\Gamma}|_{\ddot{x}^i=0}=0,
\end{equation}
whose corresponding one in E-SR is
\begin{equation}\label{38b}
\dot{\gamma}|_{\ddot{x}^i=0}\equiv{d\over dt}\left. \left({1\over
\sqrt{1-v^2/c^2}}\right)\right|_{v={\rm constant}}=0.
\end{equation}
By means of the standard procedure to perform the canonic quantization, we obtained the dS-SR wave equation for spinless  particle \cite{Yan1}:
\begin{equation}\label{KG in curved spactime}
\frac{1}{\sqrt{B}}\pa_{\mu}(B^{\mu\nu}\sqrt{B}\pa_{\nu})\phi+\frac{m_{0}^2
c^2}{\hbar^2} \phi=0,
\end{equation}
which is just the Klein-Gordon equation in  curved space-time with
Beltrami metric $B_{\mu\nu}$. The  measurable conserved 4-momentum
operator is \cite{Yan1}
\begin{eqnarray}\label{momentum operator}
p^{\mu} &=& i\hbar \left[\left(\eta^{\mu\nu}-\frac{
x^{\mu}x^{\nu}}{R^2}\right)\partial_{\nu}+\frac{5
x^{\mu}}{2R^2}\right].
\end{eqnarray}
 The
corresponding Dirac equation which describes the particle with spin
$1/2$ \cite{Ut}\cite{NY} reads
\begin{equation}\label{Cur-Dirac}
\left(ie_a^\mu \gamma^a D_\mu-{m_0c \over \hbar}\right)\psi=0,
~~{\rm or}~~\left(ie_{a\mu} \gamma^a D^\mu-{m_0c \over \hbar}\right)\psi=0,
\end{equation}
where $e_a^\mu$ is the tetrad and $D_\mu$ is the covariant
derivative with Lorentz spin connection $\omega^{ab}_\mu$. Their
definitions and relations are follows (e.g., see \cite{NY})
\begin{eqnarray}
\nonumber D_\mu &=&\pa_\mu-{i\over 4}\omega^{ab}_\mu\sigma_{ab},~~ D^\mu=B^{\mu\nu}D_\nu, \\
\nonumber \{\gamma^a, \gamma^b\}&=& 2\eta^{ab},~~\sigma_{ab}={i\over
2}[\gamma_a, \gamma_b],~~{i\over 2}[\sigma_{ab}, \sigma_{cd}]=\eta_{ac}\sigma_{bd}
-\eta_{ad}\sigma_{bc}+\eta_{bd}\sigma_{ac}-\eta_{bc}\sigma_{ad}, \\
\nonumber e_\mu^a e_\nu^b \eta_{ab}&=&B_{\mu\nu},~~e_\mu^a e_\nu^b
B^{\mu\nu}=\eta^{ab},~~e^\mu_{a\; ;\nu}=\pa_\nu e_a^\mu+ \omega_{a\;\;\nu}^{\;\;b}e_b^\mu+\Gamma^\mu_{\lambda\nu} e^\lambda_a=0,\\
\nonumber \omega^{ab}_\mu &=& {1\over 2}(e^{a\rho}\pa_\mu e^b_\rho
-e^{b\rho}\pa_\mu e^a_\rho ) -{1\over
2}\Gamma^\rho_{\lambda\mu}(e^{a\lambda}e^b_\rho
-e^{b\lambda}e^a_\rho ),\\
\label{47} \Gamma^\rho_{\lambda\mu} &=& {1\over
2}B^{\rho\nu}(\pa_\lambda B_{\nu\mu} +\pa_\mu B_{\nu\lambda}
-\pa_\nu B_{\lambda\mu}).
\end{eqnarray}
It is straightforward  to check that the components
$\psi_\alpha\;(\alpha=1,\cdots 4)$ of the spinor satisfy the
Klein-Gordon equation (\ref{KG in curved spactime}).

\section{Hydrogen atom embedded in light cone of  Friedmann-Robertson-Walker Universe}

\noindent The isotropic and homogeneous cosmology solution of Einstein equation in GR (General Relativity) is Friedmann-Robertson-Walker (FRW) metric. In this section we discuss the Hydrogen atom embedded in FRW Universe and described by dS-SR Dirac equation.

As is well known that GR can be viewed as a kind of gauge field theory \cite{Ut}\cite{Glashow}. The dynamic equation of GR can be yielded by means of the localization of external global space-time symmetries, such as Lorentz group $SL(2,C)$, transition group $T_4$, Pioncar\'{e} group etc. Such localizations make global space -time transformation $x^\mu\rightarrow\td{x}'^\mu=
\Lambda^{\mu}_{~\nu}x^\nu+a^\mu$ to be $x^\mu\rightarrow\td{x}'^\mu=
\Lambda^{\mu}_{~\nu}(x)x^\nu+a^\mu(x)=f^\mu(x)$ which is arbitrary curvilinear coordinate transformation with Christoffel symbol connection $\{^{\;\lambda}_{\mu\nu}\}$ for the torsion-free space-time, and leads to construct GR by means of Riemann tensor. It is essential that the the framework of GR is generally free to the gauge theory's underline global external space time symmetry for the torsion-free space-time. Now let's see the global de Sitter space-time transformation Eq.(\ref{transformation}). When $a^\mu\rightarrow a^\mu(x)$, $L^\mu_\nu\rightarrow L^\mu_\nu(x)$, the transformation is localizated to be $x^\mu\rightarrow \td{x}^\mu=f(x)^\mu$ which is an arbitrary curvilinear coordinate transformation. The connection is still Christoffel symbol and no corrections are yielded to the GR framework, and hence the considerations of atoms described by dS-SR QM in FRW Universe are legitimate.

One way to detect the spectrum of distance atom is  spectroscopic observations of gas clouds seen in absorption against
background Quasi-Stellar Objects (QSO), which can be used to search for level shifts of atom for various purposes
(see, e.g.,
\cite{webb06}\cite{Murphy11}).
In the observations of gas-QSO systems in the expanding Universe,
one can observe two species of frequency changes in atomic spectra:
the Hubble redshift $(z)$ caused by the usual Doppler effects and a
rest frequency change due to the dynamics of atom beyond E-SR. The latter can be
found by measuring the relative size of relativistic corrections to
the transition frequencies of atoms on the gas-QSO (or on QSO for briefness). A widely
accepted assumption is that this rest frequency change is
independent of the Hubble velocity and  cosmologic acceleration of
the gas in the Universe. Therefore all relativistic correction
calculations for atom spectra were performed in a ``rest" inertial
reference frame without any Lorentz boost and non-inertial effects
caused by frame-origin motion \cite{Dzuba1,Dzuba}.
In this present
paper, the calculations based on dS-SR Dirac equation are performed in such rest reference frame. In this
framework, the time-varying $(2s^{1/2}-2p^{1/2})$-splitting
$\Delta E(2s^{1/2}-2p^{1/2})=\Delta E(t)$ will be calculated. Furthermore, the $t-z$
relation has been established for the Universe with the observed
acceleration in $\Lambda$CDM model \cite{Lambda,Lambda1}. Employing
this relation, we have $\Delta E(t)=\Delta E(t(z))$.

 Now, we show the
dS-SR Dirac equation of Hydrogen atom on a QSO in the
Earth-QSO reference frame. As illustrated in FIG.1, the Earth is
located at the origin of frame, the proton (nucleus of Hydrogen
atom) is located at {$Q=\{Q^0\equiv c t,\;Q^1,\;Q^2=0,\;Q^3=0\}$. To
an observable atom in four-dimensional space-time, the proton has to
be located at QSO-light-cone with cosmic metric $g_{\mu\nu}$.
Namely, $Q$ must satisfy following light-cone equation (see FIG. 1
and set $Q^2=Q^3=0$ for simplification)
\begin{equation}\label{new1}
 ds^2= g_{\mu\nu}(Q)dQ^\mu
dQ^\nu=0,
\end{equation}
which determines $Q^1=f(Q^0)$.  We emphasize that the underlying
space-time symmetry for the atom near $Q$ described by
dS-SR dynamics is de Sitter group instead of to limit
it as Poincar\'{e} symmetry of E-SR as usual, which is
only a special limit of dS-SR's. The corresponding
space-time metric is
$B_{\mu\nu}(Q)=\eta_{\mu\nu}\left(1+{(Q^0)^2-(Q^1)^2\over
R^2}\right)+{1\over
  R^2}\eta_{\mu\lambda}Q^\lambda
  \eta_{\nu\rho}Q^\rho+\mathcal{O}(1/R^4)$ (see Eq.(\ref{star28})).
Note $B_{\mu\nu}(Q)$ is position $Q$-dependent, and Lorentz metric
$\eta_{\mu\nu}$ for E-SR is not. Explicitly, from
Eq.(\ref{star28}), we have
\begin{eqnarray}\label{new01}
&&\hskip-0.4in B_{\mu\nu}(Q)=\left(
\begin{array}{lccr}
1+{2(Q^0)^2-(Q^1)^2\over R^2} & -{Q^0 Q^1\over R^2} &0 &0 \\
-{Q^1 Q^0\over R^2} & -1+{2(Q^1)^2-(Q^0)^2\over R^2} & 0 & 0 \\
0 & 0 & -1-{(Q^0)^2-(Q^1)^2\over R^2} & 0 \\
0 & 0 & 0 & -1-{(Q^0)^2-(Q^1)^2\over R^2}
\end{array} \right),\\
\label{new02} &&\hskip-0.4in B^{\mu\nu}(Q)=\left(
\begin{array}{lccr}
1-{2(Q^0)^2-(Q^1)^2\over R^2} & {Q^0Q^1\over R^2} &0 &0 \\
{Q^1Q^0\over R^2} & -1-{2(Q^1)^2-(Q^0)^2\over R^2} & 0 & 0 \\
0 & 0 & -1+{(Q^0)^2-(Q^1)^2\over R^2} & 0 \\
0 & 0 & 0 & -1+{(Q^0)^2-(Q^1)^2\over R^2}
\end{array} \right).
\end{eqnarray}
In following, we will solve Eq.(\ref{new1}) to determine $Q^1$ in FRW Universe model.
 }
\begin{figure}[ht]
\begin{center}
\includegraphics[scale=0.8]{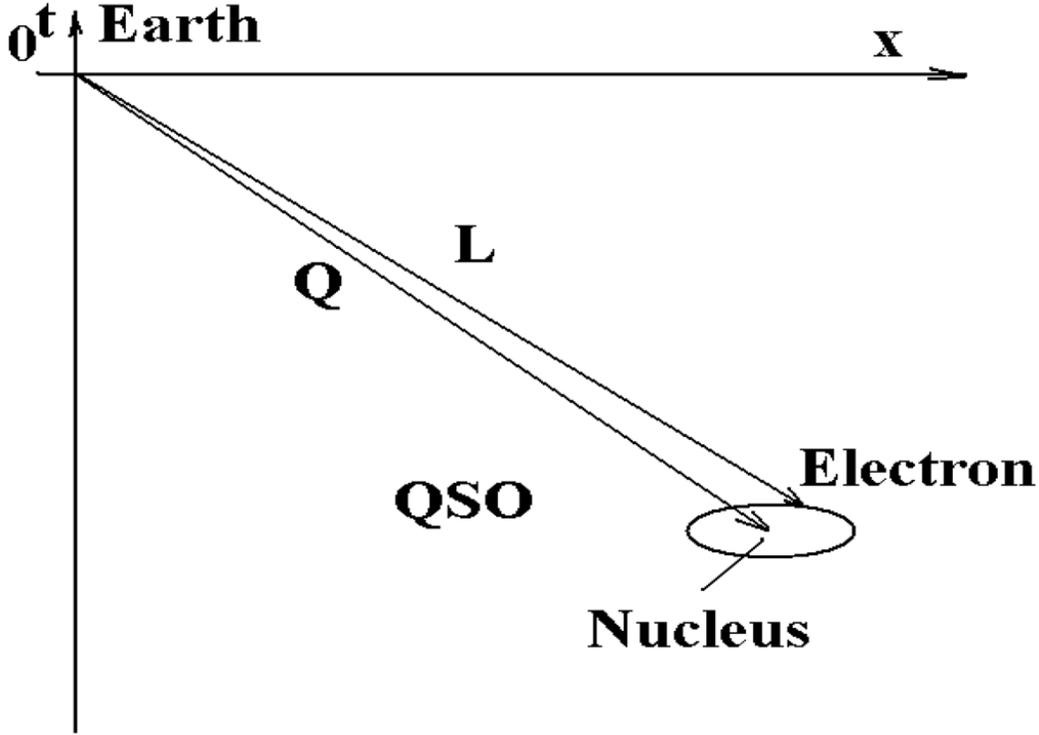}
\caption{\label{Fig1} \small Sketch of the Earth-QSO reference
frame. The Earth is located in the origin. The position vector for
nucleus of atom on QSO is $Q$, and for electron is $L$. The distance
between nucleus and electron is $r$. }
\end{center}
\end{figure}

The
Friedmann-Robertson-Walker (FRW) metric is (see, e.g., \cite{Weinberg})
\begin{eqnarray}
\nonumber ds^2&=&c^2dt^2-a(t)^2\left\{{dr^2\over
1-kr^2}+r^2d\theta^2
+r^2\sin^2\theta d\phi^2\right\} \\
\nonumber&=&(dQ^0)^2-a(t)^2\left\{dQ^idQ^i+{k (Q^idQ^i)^2\over
1-kQ^iQ^i}\right\} \\
\label{new4}&\equiv& g_{\mu\nu}(Q) dQ^\mu dQ^\nu,
\end{eqnarray}
where $r=\sqrt{Q^iQ^i},~Q^1=r\sin \theta \cos \phi,~Q^2=r\sin \theta
\sin \phi,~Q^3=r\cos\theta$ has been used. As is well know FRW
metric satisfies ¡°homogeneity and isotropy¡± principle of present
day cosmology. When $Q^2=Q^3=0$, from (\ref{new4}), we have
\begin{eqnarray} \nonumber
g_{\mu\nu}(Q)&=&\eta_{\mu\nu}-a(t)^2\delta_{\mu 1}\delta_{\nu 1}
\left( -{1\over a(t)^2}+1+{k(Q^1)^2\over 1-k(Q^1)^2}\right)\\
\label{new5}&&-(a(t)^2-1)(\delta_{\mu 2}\delta_{\nu 2}+\delta_{\mu
3}\delta_{\nu 3}).
\end{eqnarray}
For simpleness, we take $k=0$ and $a(t)=1/(1+z(t))$ (i.e., $a(t_0)=1$). And the red
shift function $z(t)$ is determined by $\Lambda$CDM model
 \cite{Lambda,Peebles,Lambda1}(see, e.g., Eq.(64) of \cite{Peebles}):
\begin{equation}\label{newLa1}
t=\int_0^z{dz' \over H(z')(1+z')},
\end{equation}
where
\begin{eqnarray}
\nonumber \label{newLa2}H(z')&=&H_0\sqrt{\Omega_{m0}(1+z')^3+\Omega_{R0}(1+z')^4+1-\Omega_{m0}},\\
\nonumber \label{newLa3}H_0&=&100\;h\simeq 100\times0.705 km\cdot s^{-1}/Mpc,\\
\label{newLa4}\Omega_{m0}&\simeq &0.274,~~~\Omega_{R0}\sim 10^{-5}.
\end{eqnarray}
Figure of $t(z)$ of Eq.(\ref{newLa1}) is shown in FIG.2.

\begin{figure}[ht]
\begin{center}
\includegraphics[scale=1.0]{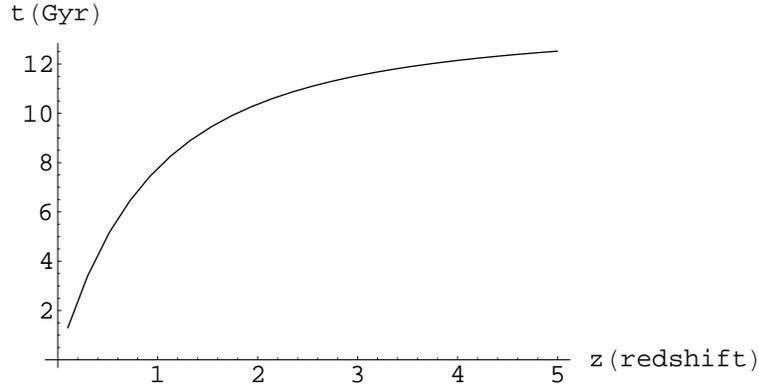}
\caption{\label{fig2}The $t-z$ relation in $\Lambda$CDM model
(eq.(\ref{newLa1})).}
\end{center}
\end{figure}

From (\ref{new5}), the FRW metric reads
\begin{eqnarray} \label{new6}
g_{\mu\nu}(Q)=\eta_{\mu\nu}-(a(t)^2-1)(\delta_{\mu 1}\delta_{\nu
1}+\delta_{\mu 2}\delta_{\nu 2}+\delta_{\mu 3}\delta_{\nu 3}).
\end{eqnarray}
Substituting (\ref{new6}) into (\ref{new1}), we have
\begin{equation}\label{new7}
dQ^0=\sqrt{-g_{11}\over g_{00}}dQ^1=a(t)dQ^1={1\over 1+z(t)}dQ^1.
\end{equation}
Consequently, by using Eq.(\ref{newLa1}) and $Q^0=c\;t$, we get
desirous result:
\begin{equation}\label{new8}
Q^1=c\int_0^z{dz'\over H(z')}.
\end{equation}
Figure of $Q^1(z)$ of Eq.(\ref{new8}) is shown in FIG.3. Ratio of $Q^1$ over $Q^0$ is shown in FIG.4.
\begin{figure}[ht]
\begin{center}
\includegraphics[scale=1.0]{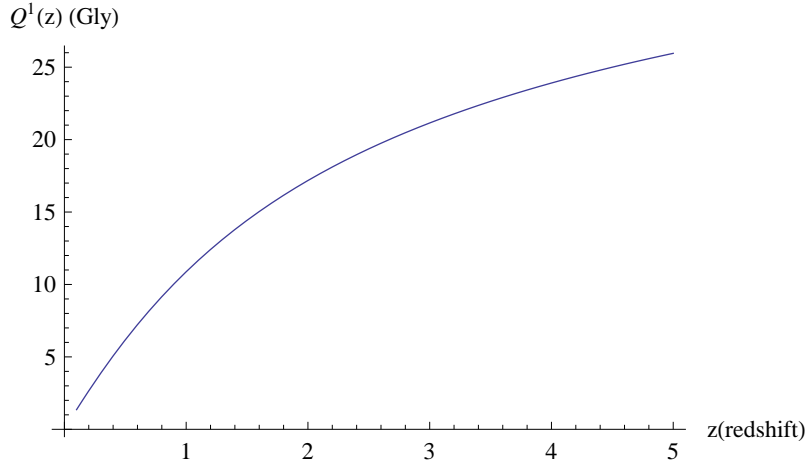}
\caption{\label{fig3}Function $Q^1(z)$ in $\Lambda$CDM model
(eq.(\ref{new8})).}
\end{center}
\end{figure}

\begin{figure}[ht]
\begin{center}
\includegraphics[scale=1.0]{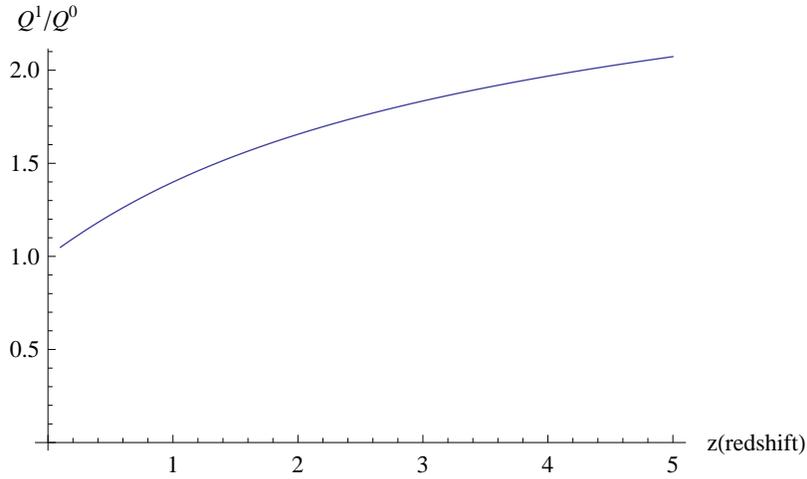}
\caption{\label{R}Function of $Q^1(z)/Q^0(z)$. $Q^1(z)$ and $Q^0(z)=ct$ are given in Eqs. (\ref{new8}) and (\ref{newLa1}).}
\end{center}
\end{figure}
 Then the location of
distance proton is  $\{Q^0, Q^1,0,0\}$ in the space-time with FRW
metric.

 We treat Hydrogen atom as a proton-electron bound state
described by quantum mechanics under instantaneous approximations
(see FIG. 1). The electron's coordinates are $L=\{L^0\equiv
ct_L\simeq ct,\;L^1,\;L^2,\;L^3\}$, and the relative space
coordinates between proton and electron are $x^i=L^i-Q^i$. The
magnitude of $r\equiv \sqrt{-\eta_{ij}x^ix^j}\sim a_B$ (where
$a_B\simeq 0.5\times 10^{-10}m$ is Bohr radius), and $|x^i|\sim
a_B$.

 According to gauge principle, the
electrodynamic interaction between the nucleus and the electron can
be taken into account by replacing the operator $D^\mu$ in
eq.(\ref{Cur-Dirac}) with the
 $U(1)$-gauge covariant derivative { $\mathcal{D}^\mu_L \equiv
D_L^\mu-ie/(c\hbar)A^\mu$, where  $A^\mu=\{\phi_B,\;{\mathbf A}\}$}. Hence, the
dS-SR Dirac equation for electron in Hydrogen at QSO
reads
\begin{equation}\label{Dirac1}
(ie_{\mu a} \gamma^a\mathcal{D}^\mu_L -{\mu c\over \hbar })\psi=0,
\end{equation}
where $\mu=m_e/(1+{m_e\over m_p})$ is the reduced mass of electron,
$\mathcal{D}^\mu_L={\pa\over \pa L_\mu}-{i\over
4}\omega^{ab\;\mu}\sigma_{ab}-ie/(c\hbar)A^\mu,$
$e^\mu_a$  and $\omega_\mu^{ab}$ have been given in
eqs.(\ref{star28}) (\ref{47}). For our purpose, we approximately write
$e^\mu_a$ and $\omega^{ab}_\mu$ up to $\mathcal{O}(1/R^2)$ as follows:
\begin{eqnarray}\label{e}
e^\mu_a &=&\left(1-{\eta_{cd}L^cL^d\over
2R^2}\right)\eta^\mu_a-{\eta_{ab}L^bL^\mu \over
2R^2}+\mathcal{O}(1/R^4),\\
\label{omega} \omega^{ab}_\mu &=& {1\over 2R^2}(\eta^a_\mu L^b-
\eta^b_\mu L^a )+\mathcal{O}(1/R^4).
\end{eqnarray}

In the following sections we are going to solve dS-SR-Dirac
equation for Hydrogen atom on QSO in FRW Universe model.
In this quantum system, there are two cosmologic length scales:
cosmic radius, say $R\sim 10^{12}ly$, and the distance between QSO
and the Earth, that is about $\sim c t$: say $R>c t> 10^8 ly$, and two
microcosmic length scales: the Compton wave length of electron
$a_c=\hbar/(m_e c)\simeq 0.3\times 10^{-12}m$, and Bohr radius $a_B=
\hbar^2/(m_e e^2)\simeq 0.5\times 10^{-10}m$. The calculations for
our purpose will be accurate up to $\mathcal{O}(c^2t^2/R^2)$. The
terms proportional to $\mathcal{O}(c^4t^4/R^4)$,
$\mathcal{O}(cta_c/R^2)$, $\mathcal{O}(cta_B/R^2)$ etc will be
omitted.

\section{Solution of usual E-SR Dirac
equation for hydrogen atom at QSO }

\subsection{Eigen values and eigen states}
\noindent At first, we show the solution of usual
E-SR Dirac equation in the Earth-QSO reference frame of
Fig.1, which serves as leading order of solution for the
dS-SR Dirac equation with $R\rightarrow \infty$ in
that reference frame. For the Hydrogen, { $\pa^\mu\rightarrow
\mathcal{D}_L^\mu=\pa_L^\mu-ie/(c\hbar)A^\mu_M$ (noting
$\omega^{ab\;\mu}|_{R\rightarrow \infty}=0$), where $A^\mu_M\equiv
\{\phi_M(x),\;{\mathbf A}_M(x)\}$, and $\phi_M(x)$ and
$\mathbf{A}_M(x)$ are nucleus electric potential and vector
potential
 at $x^i$ in Minkowski
space defined by following equations (see Appendix A, Eq.(\ref{action3}))
\begin{eqnarray}\label{potential}
-\eta^{ij}\pa_i\pa_j\phi_M(x)&=&\nabla^2\phi_M(x)=-4\pi\rho(x)=-4\pi
e\delta^{(3)}(\mathbf{x}),\\
\label{VP} \nabla(\pa_\lambda A^\lambda_M)-\pa^2{\mathbf
A}_M&=&-{4\pi\over c}{\mathbf j}=0.
\end{eqnarray}
 The solutions are $\phi_M(x)=e/r$ and $\mathbf{A}_M=0$. And hence $\pa_0\rightarrow
\mathcal{D}^L_0=\pa_0-i\eta_{00}e^2/(c\hbar r).$ }Then, the
E-SR Dirac equation reads
\begin{equation}\label{Dirac2}
i\hbar \pa_t\psi=\left(-i\hbar c \vec{\alpha} \cdot \nabla_{L}+\mu
c^2\beta -{e^2\over r}\right)\psi,
\end{equation}
where $\beta=\gamma^0,\; \alpha^i=\beta\gamma^i$. Noting the nucleus
position $\mathbf{Q}=$constant, we have
\begin{equation}\label{D3}
\nabla_{L}={\pa \over \pa \mathbf{L}} = {\pa \over \pa
(\mathbf{Q}+\mathbf{r})}={\pa \over \pa \mathbf{r}}\equiv\nabla  ,
\end{equation}
and eq.(\ref{Dirac2}) becomes the standard Dirac equation for
electron in Hydrogen at its nucleus reference frame. Energy $W$ for
E-SR-mechanics is conserved (hereafter, we use notations of \cite{Strange}), and the Hydrogen is the
stationary states of E-SR Dirac equation. The stationary
state condition is
\begin{equation}\label{S1}
i\hbar \pa_t\psi=W\psi.
\end{equation}
As is well known, combining eqs.(\ref{Dirac2}), (\ref{D3}) with
(\ref{S1}), we have
\begin{equation}\label{Dirac4}
W\psi=\left(-i\hbar c \vec{\alpha} \cdot \nabla +\mu c^2\beta
-{e^2\over r}\right)\psi\equiv H_{0}(r,\hbar,\mu,e)\psi,
\end{equation}
which is the stationary E-SR Dirac equation for
Hydrogen. The problem has been solved in terms of standard way, and
the results are follows (see, e.g., \cite{Rose}\cite{Strange})
\begin{eqnarray}\label{solution1}
W=W_{n,\kappa}&=&\mu c^2\left(1+{\alpha^2 \over
(n-|\kappa |+s)^2} \right)^{-1/2} \\ \nonumber
&&\alpha\equiv {e^2\over \hbar c},~~~~|\kappa|=(j+1/2)=1,\;2,\;3\;\cdots \\
\nonumber &&s=\sqrt{\kappa^2-\alpha^2},~~~~n=1,\;2,\;3\;\cdots.
\end{eqnarray}
And its expansion equation in $\alpha$ is
\begin{equation}\label{usual}
W=\mu c^2\left(1-{\alpha^2\over 2n^2}+\alpha^4\left({3\over
8 n^2}-{1 \over 2n^3|\kappa|}\right)+\cdots \right).
\end{equation}
The corresponding Hydrogen's wave functions $\psi$ have  also
already been finely derived (see e.g., \cite{Strange}). The complete
set of commutative observables is $\{H,\; \kappa,\;\mathbf{j}^2,\;j_z\}$,
so that $\psi=\psi_{n, \kappa, j, j_z}(\mathbf{r},\hbar, \mu, \alpha)\equiv \psi^{m_j}_j(\mathbf{r}) $,
where $\mathbf{j}=\mathbf{L}+{\hbar\over 2} \td{\sigma},\;\hbar
\kappa=\beta(\td{\sigma}\cdot \mathbf{L}+\hbar)$, and $\alpha=e^2/
(\hbar c)$. The expression of $\psi^{m_j}_j(\mathbf{r})$ is as follows
\begin{eqnarray}\label{w36}
\psi^{m_j}_j(\mathbf{r})=\left(\begin{array}{c}
g_\kappa(r)\chi^{m_j}_{\kappa}(\mathbf{\hat{r}})\\
if_\kappa(r)\chi^{m_j}_{-\kappa}(\mathbf{\hat{r}})\\
\end{array} \right)
\end{eqnarray}
where
\begin{eqnarray}\label{w37}
\chi^{m_j}_{\kappa}(\mathbf{\hat{r}})&=&\sum_{m_s=-1/2}^{1/2}C^{j\;m_j}_{l,m_j-m_s;\;1/2,m_s}Y_l^{m_j-m_s}
(\mathbf{\hat{r}})\chi^{m_s} \\
\nonumber with && \chi^{m_s=1/2}=\left(\begin{array}{c}
1\\
0\\
\end{array} \right),\;\;\;\chi^{m_s=-1/2}=\left(\begin{array}{c}
0\\
1\\
\end{array} \right)\\
\label{w38}\chi^{m_j}_{-\kappa}(\mathbf{\hat{r}})&=&-\td{\sigma}_r\chi^{m_j}_{\kappa}(\mathbf{\hat{r}}),
\;\;\; with \;\;\; \td{\sigma}_r=\mathbf{\hat{r}}\cdot \overrightarrow{\sigma}=\left(\begin{array}{cc}
\cos \theta & \sin \theta e^{-i\phi}\\
\sin \theta e^{i\phi} & -\cos \theta\\
\end{array} \right),
\end{eqnarray}
and
\begin{eqnarray}
\nonumber g_\kappa(r)&=& 2\lambda (k_C+W_C)^{1/2}e^{-\lambda r}(2\lambda r)^{s-1}\alpha_0'\\
\label{w39}&&\times \left(n'M(1-n', 2s+1,2\lambda r)+\left(\kappa-{\alpha k_C \over \lambda}\right)M(-n',2s+1,2\lambda r)\right),\\
\nonumber f_\kappa(r)&=& 2\lambda (k_C-W_C)^{1/2}e^{-\lambda r}(2\lambda r)^{s-1}\alpha_0'\\
\label{w40}&&\times \left(n'M(1-n', 2s+1,2\lambda r)-\left(\kappa-{\alpha k_C \over \lambda}\right)M(-n',2s+1,2\lambda r)\right),
\end{eqnarray}
where
\begin{eqnarray}
\nonumber && k_C=\mu c/\hbar,\;\;W_C=W/(c\hbar),\;\;\lambda=(k_C^2-W_C^2)^{1/2},\;\;s=\sqrt{\kappa^2-\alpha^2},\;\;n'=n-|\kappa|,\\
\label{notation} && \kappa=-(j(j+1)-l(l+1)+1/4),
\end{eqnarray}
where $W$ has been given in Eq.(\ref{solution1}), and $M(a,b,z)$ is the confluent hypergeometric function:
\begin{eqnarray}
\nonumber &&M(a,b,z)=1+{az\over b}+{(a)_2z^2\over 2!(b)_2}+\cdots {(a)_nz^n\over n!(b)_n}+\cdots\\
\nonumber && (a)_0=1,\;\;\;\;(a)_n=a(a+1)(a+2)\cdots(a+n-1),
\end{eqnarray}
and $\alpha'$ is the normalization constant required by
\begin{eqnarray*}
\int_0^\infty r^2(g_\kappa^2(r)+f_\kappa^2(r))dr=1.
\end{eqnarray*}

To Hydrogen-like one electron atoms with $Z> 1$, the energy level formula  and the eigen wave functions expressions  are all the same as Eqs.(\ref{solution1})-(\ref{notation})  except $\alpha\rightarrow \xi=Z\alpha$. In FIG.5 the levels of one electron atom with $Z=92$ are shown.
\begin{figure}[ht]
\begin{center}
\includegraphics[scale=1.0]{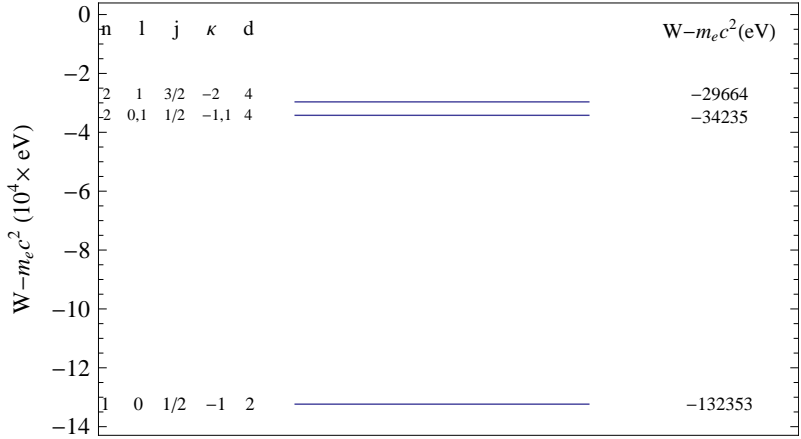}
\caption{\label{fig5}The energy levels of the one-electron atom with $Z=92$. The figure shows relativistic
energy levels calculated using equation (\ref{solution1}) with replacement of $\alpha\rightarrow \xi=Z\alpha$. They are labelled by the quantum numbers $n,\;l,\;j,\;\kappa$ and their degeneracy $d$, up to $n=2$. }
\end{center}
\end{figure}

It is learned from above that since $R\rightarrow\infty$, $B_{\mu\nu}\rightarrow\eta_{\mu\nu}$, dS-SR$\rightarrow$E-SR, the Hamiltonian of E-SR is cosmic time independent. So, the spectra of Hydrogen at any place in FRW Universe are the same, and there is almost no cosmology information of the Universe in the spectrum solutions Eq.(\ref{solution1}) of E-SR Dirac equation.

\subsection{$2s^{1/2}$ and $2p^{1/2}$ states of Hydrogen}

As is well known that the state of $2s^{1/2}$ and state of $2p^{1/2}$ are complete  degenerate to all order of $\alpha$ in the E-SR Dirac equation of Hydrogen. Namely, from (\ref{solution1}) and $\kappa=-(j(j+1)-l(l+1)+1/4)$, we have
\begin{equation}\label{2s2p}
\Delta W(2s^{1/2}-2p^{1/2})\equiv W_{(n=2,\;\kappa=-1)}-W_{(n=2,\;\kappa=+1)}=0.
\end{equation}
By means of Eqs.(\ref{w36})-(\ref{w40}), the wave functions of states $2s^{1/2}$ with $\kappa=-1$ are as follows \cite{Strange}
\begin{eqnarray}\label{w42}
\psi^{m_j}_{(2s)j=1/2}(\mathbf{r})=\left(\begin{array}{c}
g_{(2s^{1/2})}(r)\chi^{m_j}_{\kappa}(\mathbf{\hat{r}})_{(2s^{1/2})}\\
if_{(2s^{1/2})}(r)\chi^{m_j}_{-\kappa}(\mathbf{\hat{r}})_{(2s^{1/2})}\\
\end{array} \right)
\end{eqnarray}
where
\begin{eqnarray}\label{43}
&&\hskip-0.2in g_{(2s^{1/2})}(r)=\sqrt{{(2\lambda)^{2s+1}k_C (2s+1)(k_C+W_C)\over 2W_C(2W_C+k_C)\Gamma(2s+1)}}
r^{s-1}e^{-\lambda r}\left({W_C\over k_C}-{\lambda\over 2s+1}\left(1+{2W_C\over k_C}\right)r\right)\\
\label{w43}&&\hskip-0.3in f_{(2s^{1/2})}(r)=-\hskip-0.05in\sqrt{{(2\lambda)^{2s+1}k_C (2s+1)(k_C-W_C)\over 2W_C(2W_C+k_C)\Gamma(2s+1)} }
r^{s-1}e^{-\lambda r}\hskip-0.05in\left(\hskip-0.05in{W_C\over k_C}\hskip-0.05in+1
-\hskip-0.05in{\lambda\over 2s+1}\left(\hskip-0.05in 1+{2W_C\over k_C}\hskip-0.05in\right)r\right)
\end{eqnarray}
where the expressions of notations $k_C,\;W_C,\;\lambda\;$ and $s$ have been shown in Eq.(\ref{notation}) following \cite{Strange}, and due to Eqs.(\ref{w37}) (\ref{w38})
\begin{eqnarray}\label{ww46}
&&\chi^{1/2}_{\kappa}(\mathbf{\hat{r}})_{(2s^{1/2})}=\left(\begin{array}{c}
Y_0^0\\
0\\
\end{array} \right),\;
\chi^{-1/2}_{\kappa}(\mathbf{\hat{r}})_{(2s^{1/2})}=\left(\begin{array}{c}
0\\
Y_0^0\\
\end{array} \right),\;\;with\;\;Y_0^0={1\over \sqrt{4\pi}}, \\
\label{ww47} && \chi^{1/2}_{-\kappa}(\mathbf{\hat{r}})_{(2s^{1/2})}=\left(\begin{array}{c}
-\cos \theta  Y_0^0\\
-\sin\theta e^{i\phi} Y_0^0\\
\end{array} \right),\;
\chi^{-1/2}_{-\kappa}(\mathbf{\hat{r}})_{(2s^{1/2})}=\left(\begin{array}{c}
-\sin \theta e^{-i\phi} Y_0^0\\
\cos\theta Y_0^0\\
\end{array} \right).
\end{eqnarray}

For the $2p^{1/2}$ state with $\kappa'=1$
\begin{eqnarray}\label{w45}
\psi^{m_j}_{(2p)j=1/2}(\mathbf{r})=\left(\begin{array}{c}
g_{(2p^{1/2})}(r)\chi^{m_j}_{\kappa'}(\mathbf{\hat{r}})_{(2p^{1/2})}\\
if_{(2p^{1/2})}(r)\chi^{m_j}_{-\kappa'}(\mathbf{\hat{r}})_{(2p^{1/2})}\\
\end{array} \right)
\end{eqnarray}
where
\begin{eqnarray}\label{46}
&&\hskip-0.2in g_{(2p^{1/2})}(r)=\sqrt{{(2\lambda)^{2s+1}k_C (2s+1)(k_C+W_C)\over 2W_C(2W_C-k_C)\Gamma(2s+1)}}
r^{s-1}e^{-\lambda r}\hskip-0.05in\left(\hskip-0.05in{W_C\over k_C}\hskip-0.05in-1
+\hskip-0.05in{\lambda\over 2s+1}\left(\hskip-0.05in 1-{2W_C\over k_C}\hskip-0.05in\right)r\right),\\
\label{w47}&&\hskip-0.3in f_{(2p^{1/2})}(r)=-\hskip-0.05in\sqrt{{(2\lambda)^{2s+1}k_C (2s+1)(k_C-W_C)\over 2W_C(2W_C-k_C)\Gamma(2s+1)} }
r^{s-1}e^{-\lambda r}\hskip-0.05in\left({W_C\over k_C}+{\lambda\over 2s+1}\left(1-{2W_C\over k_C}\right)r\right).
\end{eqnarray}
\begin{eqnarray}\label{ww51}
&&\chi^{1/2}_{\kappa'}(\mathbf{\hat{r}})_{(2p^{1/2})}=\left(\begin{array}{c}
-\sqrt{1\over 3}Y_1^0(\theta\phi)\\
\sqrt{2\over 3}Y_1^1(\theta\phi)\\
\end{array} \right),\;
\chi^{-1/2}_{\kappa'}(\mathbf{\hat{r}})_{(2p^{1/2})}=\left(\begin{array}{c}
-\sqrt{2\over 3}Y_1^{-1}(\theta\phi)\\
\sqrt{1\over 3}Y_1^0(\theta\phi)\\
\end{array} \right),\\
\nonumber && \chi^{1/2}_{-\kappa'}(\mathbf{\hat{r}})_{(2p^{1/2})}=\left(\begin{array}{c}
\sqrt{1\over 3}\cos\theta Y_1^0(\theta\phi)-\sqrt{2\over 3}\sin\theta e^{-i\phi}Y_1^1(\theta\phi)\\
\sqrt{1\over 3}\sin\theta e^{i\phi} Y_1^0(\theta\phi)+\sqrt{2\over 3}\cos\theta Y_1^1(\theta\phi)\\
\end{array} \right),\\
\label{ww52} && \chi^{-1/2}_{-\kappa'}(\mathbf{\hat{r}})_{(2p^{1/2})}=\left(\begin{array}{c}
\sqrt{2\over 3}\cos\theta Y_1^{-1}(\theta\phi)-\sqrt{1\over 3}\sin\theta e^{-i\phi}Y_1^0(\theta\phi)\\
\sqrt{2\over 3}\sin\theta e^{i\phi} Y_1^{-1}(\theta\phi)+\sqrt{1\over 3}\cos\theta Y_1^0(\theta\phi)\\
\end{array} \right),
\end{eqnarray}
where $Y_1^0(\theta\phi)=\sqrt{3\over 4\pi}\cos\theta,\; Y_1^1(\theta\phi)=-\sqrt{3\over 8\pi}e^{i\phi}\sin\theta,\;
Y_1^{-1}(\theta\phi)=\sqrt{3\over 8\pi}e^{-i\phi}\sin\theta.$

\subsection{Hydrogen atom energy level shifts due to gravity in FRW Universe}
Secondly, we estimate the influence of external cosmological gravitational fields on the Hydrogen energy levels. 4D-FRW space-time is curved, and the atom locates in its  tangent {\it flat} space-time described by $B_{\mu\nu}|_{R\rightarrow\infty}=\eta_{\mu\nu}$. It is well known that the interaction between an external gravitational field and atom may cause some energy level shifts of the atom \cite{Parker}\cite{Moradi}.
In \cite{Moradi}, gravitational perturbations of the Hydrogen atom are derived by means of Fermi normal coordinate method \cite{Manasse}. It is found that energy level shift of one electron
atom to the first order of curvature for $nS$ states (see Eq.(8) in
\cite{Moradi}):
\begin{equation}\label{G1}
E^{(1)}(nS)={1\over
12}\zeta^{-2}\mu^{-1}n^2(5n^2+1)\mathcal{R}_{00},
\end{equation}
 where
$\mu^{-1}\zeta^{-1}=a_B$, $\zeta=e^2$ and $\mathcal{R}_{00}$ is
Reimann curvature tensor. By means of FRW metric (\ref{new6}), Einstein equation (e.g., see Eq. (1.5.18) in {\it pp.36} of \cite{Weinberg}) and $\Lambda$CDM, we have
\begin{equation}\label{G2}
\mathcal{R}_{00}=-3\ddot{a}/(c^2 a)={4\pi\over c^2}G\rho_c
(\Omega_{m0}-2\Omega_{\Lambda})\simeq -1.05\times 10^{-56}cm^{-2},
\end{equation}
and hence
\begin{eqnarray}
\nonumber E^{(1)}(nS)&=&-{1\over 6}n^2(5n^2+1)\times 7.3\times
10^{-77}eV\\
\label{G3} &=&-{1\over 6}n^2(5n^2+1)\times 1.7\times 10^{-71}(Lamb\;
shift).
\end{eqnarray}
 It is an extremely tiny level shift. To other energy levels, the corresponding level shifts are about same order magnitudes of Eq.(\ref{G3}).  Physically, it is noting but a cosmologic gravity tide effects on the atom described by E-SR Dirac equation. Eq.(\ref{G3}) indicates that such tide effects can be completely ignored. Namely the Coulomb potential in the atom is determined by Eqs.(\ref{potential}) (\ref{VP}) which arise from Eq.(\ref{action1}) of Appendix A with $g_{\mu\nu}=\eta_{\mu\nu}$, and any influences of cosmological FRW metric are able to be ignored.  Similar conclusion were reached in the de Sitter Universe model in \cite{Moradi}. In this present paper, though the atomic dynamics is of dS-SR Dirac equation rather than E-SR's, we will also ignore this sort of cosmological gravity tide effects to atoms. Namely, instead of $g_{\mu\nu}=\eta_{\mu\nu}$ for E-SR QM, we shall employe Beltromi metric $g_{\mu\nu}=B_{\mu\nu}$ to derive the Coulomb potential in the atom for dS-SR QM in the follows.

\section{dS-SR Dirac
equation for Hydrogen atom}
\noindent By eqs.(\ref{Dirac1}), (\ref{e}), (\ref{omega}),
$\pa^\mu\rightarrow \mathcal{D}^\mu_L={\pa\over \pa L_\mu}-{i\over
4}\omega^{ab\;\mu}\sigma_{ab}-ie/(c\hbar)A^\mu$, and noting
$\mathcal{D}_\mu^L=B_{\mu\nu}\mathcal{D}_L^\nu$, we have the
 dS-SR Dirac equation for the electron in Hydrogen at
 the earth-QSO reference frame as follows
\begin{equation}\label{Dirac5}
\hbar c \beta \left[i\left( 1-{\eta_{ab}L^aL^b\over
2R^2}\right)\gamma^\mu \mathcal{D}_\mu^L -{i\over 2R^2} \eta_{ab}
L^a\gamma^b L^\mu \mathcal{D}_\mu^L-{\mu c\over \hbar}
\right]\psi=0,
\end{equation}
where factor $\hbar c \beta$ in the front of the equation is only
for convenience. We expand each terms of (\ref{Dirac5}) in order as
follows:
\begin{enumerate}
\item Since observed QSO must be located on the light cone, then
$\eta_{ab}L^aL^b\simeq \eta_{ab}Q^aQ^b=(Q^0)^2-(Q^1)^2$, and the
first term of (\ref{Dirac5}) reads {
\begin{eqnarray}
\label{first01} &&\hbar c \beta i\left( 1-{\eta_{ab}L^aL^b\over
2R^2}\right)\gamma^\mu \mathcal{D}_\mu^L =
\left(1-{(Q^0)^2-(Q^1)^2\over 2R^2}\right)
\hbar c \beta i \gamma^\mu \mathcal{D}^L_\mu \psi\\
\label{first02}&&{\rm with}~~\beta\gamma^\mu=\{\beta \gamma^0=\beta^2=1,\;\beta\gamma^i=\alpha^i\}\\
\label{first} &&\hbar c \beta i \gamma^\mu \mathcal{D}^L_\mu
\psi=(i\hbar \pa_t + \hbar c \vec{\alpha}\cdot
\nabla + {\hbar c \beta \over 4}
\omega^{ab}_\mu\gamma^\mu\sigma_{ab} +e \beta\gamma^\mu
B_{\mu\nu}A^\nu ) \psi,
\end{eqnarray}
where $A^\nu=\{A^0=\phi_B,\; A^i\}$ are determined by Maxwell
equations within constance metric $g_{\mu\nu}=B_{\mu\nu}(Q)$ and
$j^\nu=\{j^0\equiv c\rho/\sqrt{B_{00}},\; j^i=0\}$ (see Appendix A):
\begin{eqnarray}
\nonumber -B^{ij}(Q)\pa_i\pa_j\phi_B(x)\hskip-0.07in
&=&\hskip-0.06in\left[\left(1-{(Q^0)^2-(Q^1)^2\over
R^2}\right)\nabla^2+{(Q^1)^2\over R^2}{\pa^2\over \pa
(x^1)^2}\right]\hskip-0.06in
\phi_B(x)= -{4\pi\over c} j^0\\
\label{potential-B} &=&{-4\pi e\over
\sqrt{-\det(B_{ij}(Q))B_{00}(Q)}}\delta^{(3)}(\mathbf{x}),
\end{eqnarray}
\begin{equation}\label{potential-BB1}
\pa^i\pa_\mu A^\mu-B^{\mu\nu}\pa_\mu\pa_\nu A^i=-{4\pi\over c}j^i=
0.
\end{equation}
 The solution is (see Appendix A)
\begin{equation}\label{potential-B1}
\phi_B=\left(1-{3(Q^0)^2-2(Q^1)^2\over 2R^2}\right){e\over
r_B},~~~A^i=0,
\end{equation}
where $r_B=\sqrt{(\td{x}^1)^2+(x^2)^2+(x^3)^2}$ with $\td{x}^1\equiv
x^1/[1+{(Q^1)^2\over 2R^2}]$. } In the follows, we use variables
$\{\td{x}^1, x^2, x^3\}$ to replace $\{x^1, x^2, x^3\}$. Following
notations are introduced hereafter:
\begin{eqnarray}\label{nota1}
\mathbf{r}_B&=&\mathbf{i}\td{x}^1+ \mathbf{j}x^2+\mathbf{k}x^3,~~~~|\mathbf{r}_B|=r_B, \\
\label{nota2} \nabla_B &=& \mathbf{i}{\pa \over \pa\td{x}^1}+
\mathbf{j}{\pa \over \pa x^2}+\mathbf{k}{\pa\over \pa
x^3},~~~~~\td{x}^i\in\{\td{x}^1,\;x^2,\;x^3\}.
\end{eqnarray}
Then, noting ${\pa\over \pa x^1}={\pa \td{x}^1\over \pa
x^1}{\pa\over \pa \td{x}^1}={\pa\over \pa \td{x}^1}-{(Q^1)^2\over
2R^2}{\pa\over \pa \td{x}^1}$, the eq.(\ref{first}) becomes
\begin{eqnarray}
\nonumber \hbar c \beta i \gamma^\mu \mathcal{D}^L_\mu \psi
&=&\left(i\hbar \pa_t +i\hbar c \vec{\alpha}\cdot \nabla_B-i\hbar
c{(Q^1)^2\over 2R^2}\alpha^1{\pa\over \pa\td{x}^1}\right.\\
\nonumber && + \left.{\hbar c \beta \over 4}
\omega^{ab}_\mu\gamma^\mu\sigma_{ab}
 +eB_{00}\phi_B+e\alpha^1 B_{10} \phi_B \right) \psi\\
\nonumber &=&\left(i\hbar \pa_t +i\hbar c \vec{\alpha}\cdot
\nabla_B-i\hbar c{(Q^1)^2\over 2R^2}\alpha^1{\pa\over \pa\td{x}^1} +
{\hbar c \beta \over 4}
\omega^{ab}_\mu\gamma^\mu\sigma_{ab}\right.\\
\label{first1} && \left. +\left[1+{2(Q^0)^2-(Q^1)^2\over R^2}\right]e\phi_B-{Q^1Q^0\over R^2}\alpha^1e\phi_B\right)  \psi.
\end{eqnarray}

\item Estimation of the contributions of the fourth term in RSH of (\ref{first1})
( the spin-connection contributions) is as follows: By (\ref{omega}), the ratio of
the fourth term to the first term of (\ref{first1}) is:
\begin{equation}\label{61}
\left|{ {\hbar c \beta \over 4}
\omega^{ab}_\mu\gamma^\mu\sigma_{ab}\psi \over i\hbar \pa_t\psi }
\right|\sim {\hbar c\over 4}{ct\over 2R^2}{1\over m_ec^2}={ct\over
8R^2}{\hbar\over m_ec}={1\over 8}{cta_c\over R^2}\sim 0,
\end{equation}
where $a_c= \hbar/(m_e c)\simeq 0.3\times 10^{-12}m$ is the Compton
wave length of electron. $\mathcal{O}(cta_c/R^2)$-term is
neglectable. Therefore the 3-rd term in RSH of (\ref{first}) has no
contribution to our approximation calculations.

\item Substituting  (\ref{61})
(\ref{first1}) (\ref{first02}) into (\ref{first01}) and noting
$L^a\simeq Q^a$ (see FIG. 1), we get the first term in LHS of
(\ref{Dirac5})
\begin{eqnarray}\nonumber
\left.\right.\hskip-0.5in \hbar c \beta i\left(
1-{\eta_{ab}L^aL^b\over 2R^2}\right) &&\hskip-0.2in \gamma^\mu
\mathcal{D}^L_\mu \psi=\left(1-{(Q^0)^2-(Q^1)^2\over
2R^2}\right)\left(i\hbar {\pa\over
\pa t} +i\hbar c \vec{\alpha}\cdot \nabla_B\right. \\
\nonumber && \hskip-0.3in \left. -i\hbar c{(Q^1)^2\over
2R^2}\alpha^1{\pa\over \pa\td{x}^1}
+[1+{2(Q^0)^2-(Q^1)^2\over R^2}]e\phi_B -{Q^1Q^0\over R^2}\alpha^1e\phi_B\right) \psi \\
\nonumber &&\hskip-0.8in =\left\{\left(1-{(Q^0)^2-(Q^1)^2\over
2R^2}\right)\left(i\hbar {\pa\over \pa t} +i\hbar c
\vec{\alpha}\cdot \nabla_B\right)-i\hbar c{(Q^1)^2\over
2R^2}\alpha^1{\pa\over \pa\td{x}^1}\right.\\
\label{first-1} && \left. +\left[1+{3(Q^0)^2-(Q^1)^2\over
2R^2}\right]e\phi_B -{Q^1Q^0\over R^2}\alpha^1e\phi_B+ \mathcal{O}(1/R^4)\right\} \psi.
\end{eqnarray}

\item The second term of (\ref{Dirac5}) is
\begin{eqnarray}\label{second}
 \nonumber -\hbar c \beta {i\over 2R^2} \eta_{ab} L^a\gamma^b L^\mu
\mathcal{D}^L_\mu \psi&=&-{i\hbar c \beta \over 2R^2}(\gamma^0
L^0-\vec{\gamma}\cdot \vec{L})L^\mu \left[\pa_\mu^L-\delta_{\mu
0}{ie\over c\hbar }\phi_B \right]\psi +\mathcal{O}({1\over R^4})
\\ \nonumber &=&-{i\hbar c \over
2R^2}(L^0-\vec{L}\cdot\vec{\alpha})\left( L^0\pa_0^L-L^0{ie\over
c\hbar}\phi_B+L^i\pa_i^L\right)\psi \\
\nonumber &\simeq& -{ic\hbar \over
2R^2}(L^0-L^1\alpha^1)(L^0\pa_0^L-L^0{ie\over
c\hbar }\phi_B+L^1\pa^L_1)\psi \\
\nonumber &\simeq& -{ic\hbar \over 2R^2}\left[\left((L^0)^2-L^1L^0\alpha^1\right)\pa_0-{ie\over c\hbar}(L^0)^2\phi_B \right.\\
&&\left.+{ie\over c\hbar}L^0L^1\alpha^1\phi_B +L^0L^1\pa^L_1-(L^1)^2\alpha^1\pa_1^L\right]\psi,
\end{eqnarray}
where following approximation estimations were used
\begin{equation}\label{note1}
{(L^2)\over R}\sim {(L^3)\over R}\sim {a_B\over R}\sim 0.
\end{equation}
Noting $L^0\simeq Q^0,\;L^1\simeq Q^1$, (\ref{second}) becomes
\begin{eqnarray}\label{second1}
 \nonumber -\hbar c \beta {i\over 2R^2} \eta_{ab} L^a\gamma^b L^\mu
\mathcal{D}^L_\mu \psi&=& \left[\left({-(Q^0)^2\over 2R^2}+{Q^1Q^0\over 2R^2}\alpha^1\right)i\hbar\pa_t-{(Q^0)^2\over 2R^2}e\phi_B+{Q^1Q^0\over 2R^2}\alpha^1e\phi_B\right. \\
&&\left.-{Q^1Q^0\over 2R^2}i\hbar c{\pa\over \pa \td{x}^1}
+{(Q^1)^2\over 2R^2}i\hbar c \alpha^1{\pa\over \pa \td{x}^1}\right]\psi.
\end{eqnarray}

\item Therefore, substituting (\ref{first-1}) (\ref{second1}) into (\ref{Dirac5}), we
have
\begin{eqnarray}\nonumber
i\hbar\left(1-{2(Q^0)^2-(Q^1)^2\over 2R^2}+{Q^1Q^0\over 2R^2}\alpha^1\right)\pa_t\psi&=&\left[
-i\hbar c \left(1-{(Q^0)^2-(Q^1)^2\over 2R^2}\right)
\vec{\alpha}\cdot\nabla_B + \mu c^2\beta \right.\\
 \label{Dirac6}&&\hskip-1.3in-
\left.\left(1+{2(Q^0)^2-(Q^1)^2\over 2R^2}\right)e\phi_B+{Q^1Q^0\over 2R^2}i\hbar c{\pa\over \pa\td{x}^1}\right]\psi,
\end{eqnarray}
or
\begin{eqnarray}\nonumber
i\hbar\pa_t\psi&=&\left[ -\left(1+{(Q^0)^2\over 2R^2}\right)i\hbar c
\vec{\alpha}\cdot\nabla_B +\left(1+{2(Q^0)^2-(Q^1)^2\over 2R^2}
\right)\mu c^2\beta
\right.\\
\nonumber &&-\left.\left( 1+{(Q^0)^2\over 2R^2}\right)
{e^2\over r_B}\right]\psi-{Q^1Q^0\over 2R^2}\alpha^1\left[-i\hbar c\overrightarrow{\alpha}\cdot \nabla_B
+\mu c^2\beta-{e^2\over r_B}\right]\psi \\
\label{rrDirac6} &&+\left[{Q^1Q^0\over 2R^2}i\hbar c{\pa\over \pa\td{x}^1}\right]\psi,
\end{eqnarray}
where $\phi_B=(1-{3(Q^0)^2-2(Q^1)^2\over
2R^2}){e/ r_B}$ were
 used (see Eq.(\ref{potential-B1})).
  Eq.(\ref{rrDirac6}) is dS-SR Dirac
 wave equation to the first order of $\mathcal{O}({c^2t^2\over
 R^2})$. Two remarks on (\ref{rrDirac6}) are as follows:

 {\bf i)} When $R\rightarrow \infty$,
  Eq.(\ref{rrDirac6}) goes back
 to usual E-SR Dirac equation of Hydrogen, which has been discussed in the last section.

{\bf ii)} Eq.(\ref{rrDirac6}) is a time-dependent wave equation.
 It is somewhat difficult to deal with the time-dependent
 problems in quantum mechanics. Generally, there are two
approximative approaches to discuss two extreme cases respectively:
(i) The modification in states obtained by the wave equation depends
critically on the time $T$ during which the modification of the
system's ``Hamiltonian" take place. For this case, one would use the
sudden approach; And, (ii), for case that of a very slow
modification of Hamiltonian,  the adiabatic approach works
\cite{Messian}. To wave equation of (\ref{rrDirac6}), like the
discussions in Introduction of this paper, since $|R|$ is
cosmologically large and $|R|>>ct$, factor
$\{(Q^0)^2/R^2,\;(Q^1)^2/R^2\}\propto (c^2t^2/ R^2)$ makes the
time-evolution of the system is so slow that the adiabatic
approximation \cite{Born} may legitimately
 works. In the below (the subsection {\bf E}), we will provide a calculations to
 confirm this point.

\end{enumerate}

\section{dS-SR spectra equation of hydrogen}

\noindent  In order to discuss the spectra of Hydrogen by
dS-SR Dirac equation, we need to find out its
solutions with certain physics energy $E$. By eq.(\ref{momentum
operator}), and being similar to
 (\ref{S1}), the dS-SR-energy eigen-state condition for
 (\ref{rrDirac6}) can be derived by means of the operator expression of momentum
 in dS-SR (\ref{momentum operator}):
\begin{eqnarray}\label{S2}
\nonumber p^0={E\over c}&=&i\hbar\left[{1\over c}\pa_t-{ct\over
R^2}x^\nu\pa^L_\nu+{5ct\over 2R^2}\right]\\
\nonumber E&=&i\hbar \left[\pa_t-{c^2t^2\over
R^2}\pa_t+{5ct\over 2R^2}\right]\\
E\psi&\simeq &i\hbar\left(1-{c^2t^2\over
R^2}\right)\pa_t\psi=i\hbar\left(1-{(Q^0)^2\over
R^2}\right)\pa_t\psi,
\end{eqnarray}
where a estimation for the ratio of the 3-rd term to the 2-nd of
$E\psi$ were used:
$${|i\hbar {5c^2t\over 2R^2}\psi|\over |{-c^2t^2\over R^2}i\hbar
\pa_t\psi|}\sim {|i\hbar {5c^2t\over 2R^2}|\over |{-2c^2t^2\over
R^2}E|}\sim {5\hbar \over 2t m_e c^2}\equiv {5\over 2}{a_c\over ct}
$$  where $a_c\simeq 0.3\times 10^{-12}$m is the Compton wave length of electron and $ct$ is about the
distance between earth and QSO. In our approximative calculations
$a_c/(ct)$ is neglectable. For instance, to a QSO with $ct\sim
10^9$ly, $a_c/(ct)\sim 10^{-38}<< (ct)^2/R^2\sim 10^{-5}$. Hence the
3-rd term of $E\psi$ were ignored.

Inserting (\ref{S2}) into (\ref{rrDirac6}), we obtain the
$\mathcal{SR}_{cR}$-spectra equation of hydrogen
\begin{eqnarray*}
\left(1+{(Q^0)^2\over R^2}\right)E\psi &=&\left[ -\left(1+{(Q^0)^2\over 2R^2}\right)i\hbar c
\vec{\alpha}\cdot\nabla_B +\left(1+{2(Q^0)^2-(Q^1)^2\over 2R^2}
\right)\mu c^2\beta
\right.\\
\nonumber &&-\left.\left( 1+{(Q^0)^2\over 2R^2}\right)
{e^2\over r_B}\right]\psi-{Q^1Q^0\over 2R^2}\alpha^1\left[-i\hbar c\overrightarrow{\alpha}\cdot \nabla_B
+\mu c^2\beta-{e^2\over r_B}\right]\psi \\
\label{rDirac6} &&+\left[{Q^1Q^0\over 2R^2}i\hbar c{\pa\over \pa\td{x}^1}\right]\psi,
\end{eqnarray*}
or
\begin{eqnarray}
\nonumber E\psi&=&\left[ -\left(1-{(Q^0)^2\over 2R^2}\right)i\hbar c
\vec{\alpha}\cdot\nabla_B +\left(1-{(Q^1)^2\over 2R^2}
\right)\mu c^2\beta
\right.\\
\nonumber &&-\left.\left( 1-{(Q^0)^2\over 2R^2}\right)
{e^2\over r_B}\right]\psi-{Q^1Q^0\over 2R^2}\alpha^1\left[-i\hbar c\overrightarrow{\alpha}\cdot \nabla_B
+\mu c^2\beta-{e^2\over r_B}\right]\psi \\
\nonumber &&+\left[{Q^1Q^0\over 2R^2}i\hbar c{\pa\over \pa\td{x}^1}\right]\psi\\
\nonumber &=&\left[ -i\hbar_t c
\vec{\alpha}\cdot\nabla_B +\mu_t c^2\beta-{e_t^2\over r_B}\right]\psi-{Q^1Q^0\over 2R^2}\alpha^1\left[-i\hbar c\overrightarrow{\alpha}\cdot \nabla_B
+\mu c^2\beta-{e^2\over r_B}\right]\psi \\
\nonumber &&+\left[{Q^1Q^0\over 2R^2}i\hbar c{\pa\over \pa\td{x}^1}\right]\psi\\
\label{rDirac6}&\equiv & (\;H_0(r_B, \hbar_t, \mu_t, e_t)+H'\;)\psi\equiv H_{(dS-SR)}\psi,
\end{eqnarray}
which is up to $\mathcal{O}(c^2t^2/R^2)$ (say again,
$\mathcal{O}(1/R^4),\;\mathcal{O}(cta_B/R^2),\;\mathcal{O}(cta_c/R^2)$
terms have been ignored), and where
\begin{eqnarray}\label{rDirac6-1}
&&H_0(r_B,\hbar_t, \mu_t, e_t)= -i\hbar_t c
\vec{\alpha}\cdot\nabla_B +\mu_t c^2\beta-{e_t^2\over r_B},\\
\label{rDirac6-2}&&H'={Q^1Q^0\over 2R^2}\left(-\alpha^1 H_0(r_B,\hbar_t, \mu_t, e_t)+i\hbar c{\pa\over \pa\td{x}^1}\right)
\end{eqnarray}
with
\begin{eqnarray}\label{hme1}
&&\hbar_t=\left(1-{(Q^0)^2\over 2
R^2}\right)\hbar= \left(1-{c^2t^2\over 2
R^2}\right)\hbar, \\
\label{hme2}&&\mu_t= \left(1-{(Q^1)^2\over
2R^2}\right)\mu, \\
\label{hme3}&& e_t = \left(1-{(Q^0)^2\over 4
R^2}\right)e= \left(1-{c^2t^2\over 4
R^2}\right)e,
\end{eqnarray}
and
\begin{eqnarray}\label{hme33}
\alpha_t\equiv {e_t^2\over \hbar_t
c}={e^2\over \hbar c}=\alpha.
\end{eqnarray}
Using notation in \cite{Strange}, $E_0=W$ and the unperturbed eigen-state equation is
\begin{equation}\label{ww70}
W\psi=H_0(\hbar_t, \mu_t, e_t)\psi= \left(-i\hbar_t c
\vec{\alpha}\cdot\nabla_B +\mu_t c^2\beta-{e_t^2\over r_B}\right)\psi,
\end{equation}
which  is same as (\ref{Dirac4}) except $\hbar
,\;\mu ,\;e$ being replaced by $\hbar_t,\;\mu_t,\;e_t$. However, since the time $t$ is dynamic variable in
the time-dependent Hamiltonian system, at this present stage we  do
not know whether $t$ can be approximately treated as a parameter in
the system. Hence,  we  cannot yet conclude $\hbar,\;\mu,\;e$ are time variations described by (\ref{hme1}),
(\ref{hme2}), (\ref{hme3}) at this stage. In the following section, we
pursue this subject.

\section{Adiabatic approximation solution to dS-SR-Dirac
spectra equation and time variation of physical constants}

\noindent Comparing (\ref{ww70}) with (\ref{Dirac4}), we can see
that there are three correction terms in (\ref{ww70}), which are
proportional to $ (c^2t^2/ R^2)$ and  service of
effects of dS-SR QM. Since $R>> ct$, we argue that the corrections due to the effects should be small, and the adiabatic
approach works quite well for solving this QM problem. In order to being sure on this point, we examine the corrections beyond adiabatic approximations in below by calculating them explicitly for a certain $z$. Suppose
$z\simeq 3\sim 4$, FIG.(\ref{R}) indicates $Q^1\approx 1.7Q^0=1.7\;ct$, and hence $(Q^1)^2\approx 3(Q^0)^2=3\;c^2t^2$  in (\ref{ww70}).
Rewriting this spectra equation (\ref{ww70})  in version of wave equation like
eq.(\ref{Dirac2}) via $E\Rightarrow i\hbar\pa_t$, we have
\begin{eqnarray}\label{Dirac8}
 i\hbar\pa_t\psi&= &H(t)\psi=[H_0(r_B, \hbar, \mu, {e})+H'_0(t)]\psi, \\
\label{Dirac9} {\rm where}~~~~~~~~~H_0(r_B,\hbar,\mu, {e})&=& -i\hbar c
\vec{\alpha}\cdot\nabla_B + \mu c^2\beta
-{{e}^2\over r_B}~ ~(see\; eq.(\ref{Dirac4}))\\
\label{Dirac10} H'_0(t)&=&-\left({ c^2t^2\over 2R^2}\right)
H_0(r_B, \hbar, 3\mu, e).
\end{eqnarray}
Suppose initial state of the atom is
$\psi(t=0)=\psi_s(\mathbf{r}_B,\hbar,\mu,{\alpha})$ where
$s=\{n_s,\;\kappa_s,\;\mathbf{j}_s^2,\;j_{sz}\}$, by eqs. (\ref{Dirac8})
(\ref{Dirac9}) (\ref{Dirac10}), and catching the
time-evolution effects, we have (see Chapter XVII of Vol II of
\cite{Messian}, and Appendix B)
\begin{eqnarray}
\label{wave1} \psi(t)\simeq\psi_s(\mathbf{r}_B,\hbar_t, \mu_t
,e_t)e^{-i{W_s\over \hbar}t}\hskip-0.05in + \hskip-0.05in \sum_{m\neq
s}{\dot{H}'_0(t)_{ms} \over i\hbar
\omega_{ms}^2}\left(e^{i\omega_{ms}t}-1\right)\psi_m(\mathbf{r}_B,\hbar_t, \mu_t,e_t) e^{\left(-i\int_0^t{W_m(\theta)\over
\hbar}d\theta\right)},
\end{eqnarray}
where $\psi_s(\mathbf{r}_B,\hbar_t, \mu_t
,e_t)$ is the adiabatic wave function, $\hbar_t$, $\mu_t$ $e_t$ are given in (\ref{hme1})$-$
(\ref{hme3}), and
\begin{eqnarray}
\nonumber  \dot{H}'_0(t)_{ms}|_{(m\neq s)}&=&\langle m |\dot{H}'_0(t)|
s\rangle |_{(m\neq s)}={- c^2t\over R^2}\langle m |H_0(r_B,
\hbar, 3\mu, e)|
s\rangle |_{(m\neq s)}\\
\nonumber &=&{-c^2t\over R^2}\langle m |\left(H_0(r_B,
\hbar, 3\mu, e)-H_0(r_B,\hbar, \mu, e)\right)|
s\rangle |_{(m\neq s)}\\
 \nonumber &\simeq &{-c^2t\over R^2}\langle
n_m,\;\kappa_m,\;\mathbf{j}^2_m,\;j_{mz}|2\mu c^2\beta|
n_s,\;\kappa_s,\;\mathbf{j}^2_s,\;j_{sz} \rangle
e^{-i(\omega_s-\omega_m)t} \\
\label{Hms}&=&{-2\mu c^4t\over R^2}\langle m|\beta|
s \rangle
e^{-i(\omega_s-\omega_m)t},\\
\label{o} \omega_{ms}&=&\omega_m-\omega_s,~~\omega_m={W_m\over
\hbar}.
\end{eqnarray}
Note, formula $\langle m| H_0(r,e)|s\rangle |_{m\neq s}=0$ has been
used in the calculations of (\ref{Hms}). The second term of
Right-Hand-Side (RHS) of eq. (\ref{wave1}) represents the quantum
transition amplitudes from $\psi_s$-state to $\psi_m$, which belong
to the  correction effects beyond adiabatic approximations. Now for showing the order of
magnitude of such corrections, we estimate $|\dot{H}_0'(t)_{ms}/\hbar
\omega_{ms}^2|$ for $s=1\equiv(1s^{1/2}, \kappa=-1, m_j=1/2),\;m=2\equiv(2s^{1/2}, \kappa=-1, m_j=1/2)$.
Noting $W_n\approx \mu c^2-\mu c^2\alpha^2/(2n^2)$, and the Compton wave length of electron $a_c=\hbar/(m_e c)\simeq
\hbar/(\mu c)\simeq 0.3\times 10^{-12}m $, from Eqs.(\ref{Hms})(\ref{o}) we have
\begin{eqnarray}\label{adia1}
\left|{\dot{H}'_0(t)_{21}\over \hbar \omega_{21}^2}\right|=\left|{128\over 9\alpha^4} \langle 2|\beta|1\rangle\right| {a_c\over R}
{ct\over R},~~ with~~
\beta=\left(
\begin{array}{lr}
I &0 \\
0 & -I
\end{array} \right),
\end{eqnarray}
where the state $\langle 2|$ has been given in Eqs.(\ref{w42})-(\ref{w47}) and the state $|1\rangle$ is as follows:
 \cite{Strange}
\begin{eqnarray}\label{w420}
|1\rangle=\psi^{m_j=1/2}_{(1s)j=1/2}(\mathbf{r})=\left(\begin{array}{c}
g_{(1s^{1/2})}(r)\chi^{1/2}_{\kappa}(\mathbf{\hat{r}})_{(1s^{1/2})}\\
if_{(1s^{1/2})}(r)\chi^{1/2}_{-\kappa}(\mathbf{\hat{r}})_{(1s^{1/2})}\\
\end{array} \right)
\end{eqnarray}
where
\begin{eqnarray}\label{430}
&& g_{(1s^{1/2})}(r)=\sqrt{{(2\lambda_1)^{2s+1}(k_C+W_{C1})\over 2k_C\Gamma(2s+1)}}
\;r^{s-1}e^{-\lambda_1 r}\\
\label{w430}&& f_{(1s^{1/2})}(r)=-\sqrt{{(2\lambda_1)^{2s+1}(k_C-W_{C1})\over 2k_C\Gamma(2s+1)} }
\;r^{s-1}e^{-\lambda_1 r}
\end{eqnarray}
where $W_{C1}\simeq m_e c^2(1-\alpha^2/2)/(c\hbar)$, $\lambda_1=\sqrt{k_C^2-W_{C1}^2}$ and
\begin{eqnarray}\label{ww460}
\chi^{1/2}_{\kappa}(\mathbf{\hat{r}})_{(1s^{1/2})}=\left(\begin{array}{c}
Y_0^0\\
0\\
\end{array} \right),\;\;\;
 \chi^{1/2}_{-\kappa}(\mathbf{\hat{r}})_{(1s^{1/2})}=\left(\begin{array}{c}
-\cos \theta  Y_0^0\\
-\sin\theta e^{i\phi} Y_0^0\\
\end{array} \right).
\end{eqnarray}
By means of expressions of $\psi^{m_j=1/2}_{(1s)j=1/2}(\mathbf{r})$ (Eqs.(\ref{w420})-(\ref{ww460})) and
$\psi^{m_j=1/2}_{(2s)j=1/2}(\mathbf{r})$ (Eqs.(\ref{w42})-(\ref{w47})), we have
\begin{equation}\label{beta}
\langle 2|\beta|1\rangle=\int_0^\infty dr r^2 \left(g_{(1s^{1/2})}(r)g_{(2s^{1/2})}(r)-f_{(1s^{1/2})}(r)f_{(2s^{1/2})}(r) \right)\simeq -1.12\times 10^{-5}.
\end{equation}
Substituting (\ref{beta}) into (\ref{adia1}), we obtain
\begin{equation}\label{adia2}
\left|{\dot{H}_0'(t)_{21}\over \hbar \omega_{21}^2}\right|=5.6\times 10^{4}\times {a_c\over R}
{ct\over R}
\end{equation}
Considering Compton wave length of electron $a_c={\hbar\over m_e c}\simeq 0.3\times 10^{-12}m \simeq 0.3\times 10^{-28}ly $ and both $Q^0=ct$ and $R$ are cosmological large length scales, and $R>Q^0$, therefore we have that
\begin{equation}\label{adia3}
\left|{\dot{H}'_0(t)_{21}\over \hbar \omega_{21}^2}\right|\simeq {(1.7\times 10^{-24}ly)\over R}\times{Q^0\over R}<<1.
\end{equation}
To generic $\langle m|$ and $|s\rangle$, similar to the calculations of Eq.(\ref{adia2}), we always have
\begin{equation}\label{adia4}
\left|{\dot{H}'_0(t)_{ms}\over \hbar \omega_{ms}^2}\right|\simeq (constant)\cdot {a_c\over R}\times{Q^0\over R}.
\end{equation}
Since the $(constant)$ at last is about $\sim 10^{10}$, and then $\left|{\dot{H}'_0(t)_{ms}\over \hbar \omega_{ms}^2}\right|\simeq {(1.7\times 10^{-18}ly)\over R}\times{Q^0\over R}$,  we finally obtain
\begin{equation}\label{adia5}
\left|{\dot{H}'_0(t)_{ms}\over \hbar \omega_{ms}^2}\right|<<1,
\end{equation}
which indicates the second term of adiabatic expansion expression Eq.(\ref{wave1}) can be ignored, or the corrections from beyond adiabatic approximations are quite small (or tiny), and hence the adiabatic approximation is legitimate for solving this dS-SR Dirac equation of the atom.

Thus, we achieve an interesting consequence that the fundamental
physics constants variate adiabatically along with cosmologic time
in dS-SR quantum mechanics framework. As is well
known that the quantum evolution in the time-dependent quantum
mechanics has been widely accepted and studied during past several
decades (see, e.g., \cite{Bayfield}). It is remarkable that the
time-variations of $\mu\;(or\;m_e),\;\hbar$ and $e$ (see Eqs. (\ref{hme1}) (\ref{hme2}) (\ref{hme3})) belong to  such quantum evolution
effects.

\section{$\mathbf{2s^{1/2}}$-$\mathbf{2p^{1/2}}$ splitting in the dS-SR Dirac equation of Hydrogen}

\noindent In the Section IV, we have pointed that the state of $2s^{1/2}$ and state of $2p^{1/2}$  are complete  degenerate to all order of $\alpha$ in the E-SR Dirac equation of Hydrogen described in  Hamiltonian $H_0(r,\hbar,\mu,e)$ of Eq.(\ref{Dirac4}). In this section, we calculate the $(2S^{1/2}-2p^{1/2})$-splitting duo to dS-SR effects.

\subsection{Energy levels shifts of Hydrogen in dS-SR QM as perturbation effects of E-SR Dirac equation of atom}

The dS-SR Dirac spectrum equation for Hydrogen atom has been derived in Section VI (\ref{rDirac6}), which is as follows
\begin{equation}\label{splitting1}
H_{(dS-SR)}\psi=(\;H_0(r, \hbar_t, \mu_t, e_t)+H'\;)\psi=E\psi,
\end{equation}
where
\begin{eqnarray}\label{splitting2}
&&H_0(r,\hbar_t, \mu_t, e_t)= -i\hbar_t c
\vec{\alpha}\cdot\nabla +\mu_t c^2\beta-{e_t^2\over r},\\
\label{splitting22} &&H'={1\over 2}(H'^\dag+H')\equiv H'_1+H'_2,
\end{eqnarray}
where
\begin{eqnarray}
\label{splitting3}&&H'_1 =-{Q^1Q^0\over 4R^2}\left(\alpha^1 H_0(r,\hbar_t, \mu_t, e_t)+ H_0(r,\hbar_t, \mu_t, e_t)\alpha^1\right),\\
\label{splitting4}&&H'_2 ={Q^1Q^0\over 4R^2}\left(i\hbar c\overrightarrow{{\pa\over \pa{x}^1}}-i\hbar c\overleftarrow{{\pa\over \pa{x}^1}}\right).
\end{eqnarray}
Comparing above equations with Eqs. (\ref{rDirac6})$-$(\ref{rDirac6-2}), hereafter we have removed the subscript $B$ of $r_B$ and $\nabla_B$, and the tilde notation $\sim$ of $\td{x}^1$ for simplicity. And we have also rewritten the perturbation Hamiltonian $H'$ to be explicit hermitian. In the spherical polar coordinates system the operator ${\pa\over\pa x^1}\equiv \pa_1$ in Eq.(\ref{splitting3}) is as follows
\begin{equation}\label{sp4}
\pa_1={\pa\over\pa x^1}=\overrightarrow{i}\cdot \nabla=
\sin \theta\cos\phi{\pa\over\pa r}+
\cos\theta\cos\phi{1\over r}{\pa\over\pa\theta}-{\sin\phi\over r\sin\theta}{\pa\over\pa\phi}.
\end{equation}

Comparing dS-SR QM with ordinary E-SR Dirac equation of Hydrogen, there are two distinguishing effects in the dS-SR QM descriptions of distant Hydrogen atom (or one-electron atom) in the Earth-QSO reference frame: 1, The physical constants variations with cosmic time adiabatically, which have been discussed in the previous section (see Eqs. (\ref{hme1})$-$(\ref{hme3})); 2, Perturbation effects arisen from $H'$ of Eq.(\ref{splitting22}) in $H_{(dS-SR)}$ of Eq.(\ref{splitting1}). In this section, we focuss the latter.

For adiabatic quantum system, the states are quasi-stationary in all
instants. And hence to all instants the quasi-stationary perturbation theory works. When $H_{(dS-SR)}=H_0(r,\hbar_t,\mu_t, e_t)+H'$, the unperturbed quasi-stationary solutions of $H_0(r,\hbar_t,\mu_t, e_t)\psi=W\psi$ are the same as Eqs.(\ref{Dirac4})$-$(\ref{notation}) except $\hbar\rightarrow\hbar_t,\;\mu\rightarrow \mu_t,\;e\rightarrow e_t$. Then the energy levels shifts due to $H'$  of Eq.(\ref{splitting22}) are computable in practice by the perturbation approach in QM.

Those shifts $\Delta E^i\equiv W_i-E$  due to $H'$ are determined by
\begin{equation}\label{sp5}
\det \left(\langle H_{(dS-SR)}\rangle_{ii'}- E\delta_{ii'}\right)=0,
\end{equation}
where $i=\{n,\;l,\;j,\;\kappa,\;m_j\}$, $H_{(dS-SR)}$ has been given in Eq.(\ref{splitting1}) and the elements are
\begin{equation}\label{sp6}
\langle H_{(dS-SR)}\rangle_{ii'}=\langle i| H_{0}|i'\rangle +\langle i| H'|i'\rangle
=W_{i}\delta_{ii'}+\langle  H'\rangle_{ii'},
\end{equation}
where $W_i=W_{n,\kappa}$ are shown in Eq.(\ref{solution1}).

Firstly, we  compute the elements $\langle H'\rangle_{ii'}\equiv\langle (H'_1+H'_2)\rangle_{ii'}$ with $i=i'$. From Eq.(\ref{splitting3}), we have
\begin{eqnarray}
\nonumber &&\langle H'_1\rangle_{ii}=-{Q^1Q^0\over 2R^2}W_i\langle i|\alpha^1|i \rangle \\
\nonumber &&=-{Q^1Q^0\over 2R^2}W_i\int dr r^2\int d\Omega \left(g_\kappa(r)\chi^{m_j\dag}_{\kappa}(\mathbf{\hat{r}}),
-if_\kappa(r)\chi^{m_j\dag}_{-\kappa}(\mathbf{\hat{r}})\right)
\left(\begin{array}{cc}
0 & \sigma^1\\
\sigma^1 & 0\\
\end{array} \right)
\left(\begin{array}{c}
g_\kappa(r)\chi^{m_j}_{\kappa}(\mathbf{\hat{r}})\\
if_\kappa(r)\chi^{m_j}_{-\kappa}(\mathbf{\hat{r}})\\
\end{array} \right)\\
\label{sp7}&&\propto \int d\Omega \left(\chi^{m_j\dag}_{\kappa}(\mathbf{\hat{r}})\sigma^1\chi^{m_j}_{-\kappa}(\mathbf{\hat{r}})
-\chi^{m_j\dag}_{-\kappa}(\mathbf{\hat{r}})\sigma^1\chi^{m_j}_{\kappa}(\mathbf{\hat{r}})\right).
\end{eqnarray}
Substituting Eqs.(\ref{w37}) (\ref{w38}) into Eq.(\ref{sp7}), we get
\begin{eqnarray}
\label{sp8}&& \int d\Omega \chi^{m_j\dag}_{\kappa}(\mathbf{\hat{r}})\sigma^1\chi^{m_j}_{-\kappa}(\mathbf{\hat{r}})=0\\
\label{sp9}&&\int d\Omega \chi^{m_j\dag}_{-\kappa}(\mathbf{\hat{r}})\sigma^1\chi^{m_j}_{\kappa}(\mathbf{\hat{r}})=0,
\end{eqnarray}
and hence
\begin{equation}\label{sp10}
\langle H'_1\rangle_{ii}=0.
\end{equation}
It is easy to be sure the validness of Eqs.(\ref{sp8}) (\ref{sp9}). Considering, for instance, the case of state
$|i\rangle=|(2p^{1/2})^{m_j=1/2}\rangle $ (see Eqs. (\ref{ww51}) (\ref{ww52})), we can calculate the left sides of (\ref{sp8}) (\ref{sp9}) explicitly:
\begin{eqnarray}
\nonumber &&\int d\Omega \chi^{1/2\dag}_{\kappa=1}(\mathbf{\hat{r}})_{(2p^{1/2})}\sigma^1\chi^{1/2}_{-\kappa=-1}(\mathbf{\hat{r}})_{(2p^{1/2})}\\
\nonumber &&=\int d\Omega \left(-\sqrt{1\over 3}Y_1^0(\theta\phi)^\dag ,
\sqrt{2\over 3}Y_1^1(\theta\phi)^\dag \right)\left(\begin{array}{cc}
0 & 1\\
1 & 0\\
\end{array} \right)
\left(\begin{array}{c}
\sqrt{2\over 3}\cos\theta Y_1^{-1}(\theta\phi)-\sqrt{1\over 3}\sin\theta e^{-i\phi}Y_1^0(\theta\phi)\\
\sqrt{2\over 3}\sin\theta e^{i\phi} Y_1^{-1}(\theta\phi)+\sqrt{1\over 3}\cos\theta Y_1^0(\theta\phi)\\
\end{array} \right)\\
\label{sp10}&&={-i\over 2\pi}\int_0^{2\pi}d\phi\sin \phi\int_0^\pi d\theta\sin\theta\cos^2\theta (\sin\theta-\cos\theta )=0,\\
\label{sp11} &&\int d\Omega \chi^{1/2\dag}_{-\kappa=-1}(\mathbf{\hat{r}})_{(2p^{1/2})}\sigma^1\chi^{1/2}_{\kappa=1}(\mathbf{\hat{r}})_{(2p^{1/2})}
=\left(\int d\Omega \chi^{1/2\dag}_{\kappa=1}(\mathbf{\hat{r}})_{(2p^{1/2})}\sigma^1\chi^{1/2}_{-\kappa=-1}(\mathbf{\hat{r}})_{(2p^{1/2})}
\right)^\dag=0.
\end{eqnarray}
Therefore Eqs.(\ref{sp8}) (\ref{sp9}) hold to be true for the state of
$|i\rangle=|(2p^{1/2})^{m_j=1/2}\rangle $.

From Eq.(\ref{splitting4}), we vave
\begin{eqnarray}\label{sp12}
\langle H'_2\rangle_{ii}={Q^1Q^0\over 4R^2}\langle i|\left(i\hbar c\overrightarrow{{\pa\over \pa{x}^1}}-i\hbar c\overleftarrow{{\pa\over \pa{x}^1}}\right)|i\rangle=-{Q^1Q^0\over 2R^2}c\langle i|\hat{p}^1|i\rangle=0,
\end{eqnarray}
where $\hat{p}^1=-i{\pa\over \pa x^1}$, and the fact that the average value for $p^1$ to the stationary bound state $|i\rangle$ must be vanish have been used. Eq.(\ref{sp12}) can also be checked by explicit calculations based  the known wave functions Eqs.(\ref{w36})$-$(\ref{w38}).

Combining Eq.(\ref{sp12}) with Eq.(\ref{sp10}), we find out that
\begin{eqnarray}\label{sp13}
\langle H'\rangle_{ii}=\langle H'_1\rangle_{ii}+\langle H'_2\rangle_{ii}=0,
\end{eqnarray}
which means $\langle H'\rangle_{ii'}$ is an off-diagonal matrix in the Hilbert space. As is well known that the energy shifts for non-degenerate levels due to $H'$ are as
\begin{eqnarray}\label{sp14}
\Delta E^i\equiv W_i-E=\langle H'\rangle_{ii}+\sum_{i'\neq i}{'|\langle H'\rangle_{i'i}|^2\over W_i-W_{i'}}+\cdots,
\end{eqnarray}
which could be thought as the perturbation solution of Eq.(\ref{sp5}) for non-degenerate case. Thus, since
$H'\propto \mathcal{O}(1/R^2)$ and noting Eq.(\ref{sp13}), the non-degenerate level shifts $\Delta E^i$ are $\mathcal{O}(1/R^4)$, which are beyond the considerations of this paper. We will show in next subsection that the meaningful $\mathcal{O}(1/R^2)$-energy level shifts due to off-diagonal perturbation interaction $H'$ are  occurred for degeneration levels. Typical example is $2S^{1/2}-2p^{1/2}$ splitting due to $H'$.

\subsection{$\mathbf{2s^{1/2}-2p^{1/2}}$ splitting caused by $\mathbf{H'}$}

In the Section IV, we have shown that the state of $2s^{1/2}$ and state of $2p^{1/2}$ are complete  degenerate to all order of $\alpha$ in the E-SR Dirac equation of Hydrogen (see Eq.(\ref{2s2p})). The degenerate will be broken by the effects of $H'$. In this subsection we calculate the ${2s^{1/2}-2p^{1/2}}$ splitting caused by ${H'}$.

By using the explicit expressions of $2s^{1/2}$- and $2p^{1/2}$ wave functions Eqs.(\ref{w42})$-$(\ref{ww52}), all matrix elements of $H'=H_1'+H_2'$ between them can be calculated, i.e,
\begin{eqnarray}\label{ssp1}
\langle H'\rangle_{2L^{1/2}, 2L'^{1/2}}^{m_j,\; m'_{j}}=\langle H'_1\rangle_{2L^{1/2}, 2L'^{1/2}}^{m_j,\; m'_{j}}
+\langle H'_2\rangle_{2L^{1/2}, 2L'^{1/2}}^{m_j,\; m'_{j}},
\end{eqnarray}
where
\begin{eqnarray}\nonumber
\langle H'_i\rangle_{2L^{1/2}, 2L'^{1/2}}^{m_j,\; m'_{j}}&=&\langle (2L^{1/2})^{m_j}| H'_i|(2L'^{1/2})^{m'_{j}}\rangle\\
\label{ssp1-1} &=&\int dr r^2\int d\Omega\; \psi^{m_j \dag}_{(2L)j=1/2}(\mathbf{r})\;H'_i \;\psi^{m'_j}_{(2L')j=1/2}(\mathbf{r}),
\end{eqnarray}
where $\{L, L'\}=\{s, p\}$, $i=1, 2$, and $\psi^{m_j}_{(2L)j=1/2}(\mathbf{r})$ are given in Eqs.(\ref{w42})$-$(\ref{ww52}). The matrix element calculations are presented in step by step in Appendix C, and the results are listed as follows:
\begin{enumerate}
\item $H'_1$-matrix elements: $H'_1$ is given in (\ref{splitting3}), i.e.,
\begin{eqnarray}\label{ssp2}
 H'_1 =-{Q^1Q^0\over 4R^2}\left(\alpha^1 H_0(r,\hbar, \mu, e)+ H_0(r,\hbar, \mu, e)\alpha^1\right),
\end{eqnarray}
where the subscript $t$ of $\hbar, \mu, e$ has been removed, because there is already a factor of $(1/R^2)$ in the $H'_1$-expression and $1/R^4$-terms are ignorable. By means of straightforward calculations (see Appendix C), we have
\begin{eqnarray}
\nonumber && \langle H'_1\rangle_{2s^{1/2}, 2p^{1/2}}^{1/2,\; -1/2}=\langle H'_1\rangle_{2s^{1/2}, 2p^{1/2}}^{-1/2,\; 1/2}
=-\langle H'_1\rangle_{2p^{1/2}, 2s^{1/2}}^{1/2,\; -1/2}=-\langle H'_1\rangle_{2p^{1/2}, 2s^{1/2}}^{-1/2,\; 1/2}\\   \label{ssp3}&&=-i{Q^1Q^0\over 4R^2}{W\over3}\sqrt{k_C^2-W_C^2\over 4W_C^2-k_C^2}\left({W_C\over k_C}-{k_C\over 2W_C}(s+1)\right)\equiv -i\Theta_1,\\
\nonumber &&\\
\nonumber &&{\rm and\; others}=0,
\end{eqnarray}
where $W=W_{(n=2, \kappa=\pm 1)}$ (see Eq.(\ref{solution1})), and notations of $k_C, W_C, \lambda, s, \kappa $ have been given in Eq.(\ref{notation}).
The above results can be expressed in explicit matrix form as follows
\begin{eqnarray}\nonumber
&&\hskip0.6in (2L^j,L^{'j'})^{m_j,m'_j}:(2s^{1/2})^{1/2}~~(2s^{1/2})^{-1/2}~~(2p^{1/2})^{1/2}~~~(2p^{1/2})^{-1/2}\\
\label{ssp4}&& \{\langle H'_1\rangle_{ii'}\} =\hskip0.7in\left(
 \begin{array}{ccccc}
 \hskip-0.9in (2s^{1/2})^{1/2} & \hskip0.2in 0 &\hskip0.4in 0 &\hskip0.4in 0 &\hskip0.4in -i\Theta_1 \\
\hskip-0.9in (2s^{1/2})^{-1/2} & \hskip0.2in 0 &\hskip0.4in 0  &\hskip0.4in -i\Theta_1 &\hskip0.4in 0 \\
\hskip-0.9in (2p^{1/2})^{1/2}  &\hskip0.2in 0  &\hskip0.4in i\Theta_1 &\hskip0.4in 0 & \hskip0.4in 0\\
\hskip-0.9in (2p^{1/2})^{-1/2}  &\hskip0.2in i\Theta_1 &\hskip0.4in 0 & \hskip0.4in 0 &\hskip0.4in 0
 \end{array}\right),
\end{eqnarray}
where the matrix row's and column's indices have been labeled explicitly. Compactly, Eq.(\ref{ssp4}) can be written as follows
\begin{eqnarray}\label{ssp5}
&&\{\langle H'_1\rangle_{ii'}\}=\left(
\begin{array}{cc}
0 & -i\Theta_1\sigma^1\\
i\Theta_1\sigma^1 & 0
\end{array}\right),~~{\rm with}~~\sigma^1=\left( \begin{array}{cc}
                                                    0 & 1 \\
                                                    1 & 0
                                                  \end{array}\right),\\
\label{ssp6}&& \hskip0.2in {\rm and}~\Theta_1={Q^1Q^0\over 4R^2}{W\over3}\sqrt{k_C^2-W_C^2\over 4W_C^2-k_C^2}\left({W_C\over k_C}-{k_C\over 2W_C}(s+1)\right).
\end{eqnarray}
\item $H'_2$ matrix elements: $H'_2$ is shown in Eq.(\ref{splitting4}):
\begin{eqnarray}\label{ssp7}
&&H'_2 ={Q^1Q^0\over 4R^2}\left(i\hbar c\overrightarrow{{\pa\over \pa{x}^1}}-i\hbar c\overleftarrow{{\pa\over \pa{x}^1}}\right),\\
\label{ssp8}&&{\rm where}~~ {\pa\over\pa x^1}=
\sin \theta\cos\phi{\pa\over\pa r}+
\cos\theta\cos\phi{1\over r}{\pa\over\pa\theta}-{\sin\phi\over r\sin\theta}{\pa\over\pa\phi}.
\end{eqnarray}
By means of straightforward calculations (see Appendix C), we obtain all elements of $H'_2$:
\begin{eqnarray}
\nonumber && \langle H'_2\rangle_{2s^{1/2}, 2p^{1/2}}^{1/2,\; -1/2}=\langle H'_2\rangle_{2s^{1/2}, 2p^{1/2}}^{-1/2,\; 1/2}
=-\langle H'_2\rangle_{2p^{1/2}, 2s^{1/2}}^{1/2,\; -1/2}=-\langle H'_2\rangle_{2p^{1/2}, 2s^{1/2}}^{-1/2,\; 1/2}\\   \label{ssp9}&&=i{Q^1Q^0\over 2R^2}{\hbar c\lambda \over 6\sqrt{4W_C^2-k_C^2}}\left({k_C^2\over W_C}-2({1\over s}+1)k_C-W_C+{2\over k_C s}W^2_C\right)\equiv -i\Theta_2,\\
\nonumber &&\\
\nonumber &&{\rm and\; others}=0.
\end{eqnarray}
The matrix form of (\ref{ssp9}) is as follows:
\begin{eqnarray}\label{ssp10}
\{\langle H'_2\rangle_{ii'}\}=\left(
\begin{array}{cc}
0 & -i\Theta_2\sigma^1\\
i\Theta_2\sigma^1 & 0
\end{array}\right),
\end{eqnarray}
where
\begin{eqnarray}
\label{ssp11}\Theta_2=-{Q^1Q^0\over 2R^2}{\hbar c\lambda \over 6\sqrt{4W_C^2-k_C^2}}\left({k_C^2\over W_C}-2({1\over s}+1)k_C-W_C+{2\over k_C s}W^2_C\right).
\end{eqnarray}

\item Since $H'=H'_1+H'_2$ (see Eq.(\ref{splitting22})), the matrix form of $H'$ is:
\begin{eqnarray}\label{ssp12}
\{\langle H'\rangle_{ii'}\}=\{\langle H'_1\rangle_{ii'}\}+\{\langle H'_2\rangle_{ii'}\}=\left(
\begin{array}{cc}
0 & -i\Theta\sigma^1\\
i\Theta\sigma^1 & 0
\end{array}\right),
\end{eqnarray}
where
\begin{eqnarray}
\nonumber \Theta=\Theta_1+\Theta_2&=&{Q^1Q^0\over 4R^2}{W\over3}\sqrt{k_C^2-W_C^2\over 4W_C^2-k_C^2}\left({W_C\over k_C}-{k_C\over 2W_C}(s+1)\right)  \\
\label{ssp13} &&\hskip-0.2in-{Q^1Q^0\over 2R^2}{\hbar c\lambda \over 6\sqrt{4W_C^2-k_C^2}}\left({k_C^2\over W_C}-2({1\over s}+1)k_C-W_C+{2\over k_C s}W^2_C\right).
\end{eqnarray}
Obviously, $\Theta\propto \mathcal{O}(1/R^2)$.

\item Substituting Eq.(\ref{ssp12}) into Eq.(\ref{sp6}) and Eq.(\ref{sp5}),  we get the secular equation for eigen value $E$ in the degenerate perturbation calculations:
\begin{eqnarray}
\label{ssp14}\left|
 \begin{array}{cccc}
 W-E & 0 & 0 & -i\Theta \\
 0 &W-E  & -i\Theta & 0 \\
 0  & i\Theta &W-E & 0\\
 i\Theta & 0 &  0 & W-E
 \end{array}\right|=0.
\end{eqnarray}
The real energy solutions are
\begin{eqnarray}\label{ssp15}
W-E=\pm\Theta,~~{\rm or}~~E^{(+)}=W+\Theta,~~E^{(-)}=W-\Theta,
\end{eqnarray}
and hence
\begin{eqnarray}
\nonumber &&(\Delta E)_{(2s^{1/2}-2p^{1/2})}\equiv E^{(+)}-E^{(-)}=2\Theta \\
\nonumber &&={Q^1Q^0\over R^2}\left[{W\over6}\sqrt{k_C^2-W_C^2\over 4W_C^2-k_C^2}\left({W_C\over k_C}-{k_C\over 2W_C}(s+1)\right)\right.\\
\label{ssp16} &&\left.\hskip0.2in  -{\hbar c\lambda \over 6\sqrt{4W_C^2-k_C^2}}\left({k_C^2\over W_C}-2({1\over s}+1)k_C-W_C+{2\over k_C s}W^2_C\right)\right],
\end{eqnarray}
which is $\mathcal{O}(1/R^2)$ and the desired expression of energy level splitting of $(2s^{1/2}-2p^{1/2})$ due to $H'$. Eq.(\ref{ssp16}) represents an important effect of dS-SR to one-electron atom, and it is the main result of this paper.

\item The eigenstates with eigenvalues $E^{(\pm)}$: Generally, the eigenstates of $H'$ are
\begin{eqnarray}\label{ssp17}
|E^{(\pm)}\rangle=C_1^{(\pm)}|2s^{1/2}\rangle^{1/2}+C_2^{(\pm)}|2s^{1/2}\rangle^{-1/2}+
C_3^{(\pm)}|2p^{1/2}\rangle^{1/2}+C_4^{(\pm)}|2p^{1/2}\rangle^{-1/2},
\end{eqnarray}
where $\{|2s^{1/2}\rangle^{m_j},\;|2s^{1/2}\rangle^{m_j}\}\in |nL^{j}\rangle^{m_j}$, and $C_i^{(\pm)}\;(i=1,2,3,4)$ satisfy the following eigen-equation corresponding to Eq.(\ref{ssp14}):
\begin{eqnarray}
\label{ssp18}\left(
 \begin{array}{cccc}
 W & 0 & 0 & -i\Theta \\
 0 &W  & -i\Theta & 0 \\
 0  & i\Theta &W & 0\\
 i\Theta & 0 &  0 & W
 \end{array}\right) \left(\begin{array}{c}
                          C_1^{(\pm)}\\
                          C_2^{(\pm)}\\
                          C_3^{(\pm)}\\
                          C_4^{(\pm)}\end{array}\right)
 =E^{(\pm)}\left(\begin{array}{c}
                          C_1^{(\pm)}\\
                          C_2^{(\pm)}\\
                          C_3^{(\pm)}\\
                          C_4^{(\pm)}\end{array}\right).
\end{eqnarray}
Substituting (\ref{ssp15}) into (\ref{ssp18}), we get the eigenstates with eigenvalues $E^{(\pm)}$:
\begin{eqnarray}\label{ssp19}
&&|E^{(+)}\rangle={1\over2}(|2s^{1/2}\rangle^{1/2}+|2s^{1/2}\rangle^{-1/2}+
i|2p^{1/2}\rangle^{1/2}+i|2p^{1/2}\rangle^{-1/2}),\\
\label{20}&&|E^{(-)}\rangle={1\over2}(|2s^{1/2}\rangle^{1/2}+|2s^{1/2}\rangle^{-1/2}-
i|2p^{1/2}\rangle^{1/2}-i|2p^{1/2}\rangle^{-1/2}),
\end{eqnarray}
which satisfy $\langle E^{(+)}|E^{(+)}\rangle=\langle E^{(-)}|E^{(-)}\rangle=1,\;\rm{and}\;\langle E^{(-)}|E^{(+)}\rangle=0$.

\item Numerical discussions: In order to getting comprehensive indications to Eq.(\ref{ssp16}), we now discuss
$(\Delta E)_{(2s^{1/2}-2p^{1/2})}\equiv E^{(+)}-E^{(-)}=2\Theta$ numerically. To Hydrogen's $2s^{1/2}$- and $2p^{1/2}$-states, $\mu\simeq m_e= 510998.910\;eV/c^2$, $Z=1,\;n=2,\;\kappa=\pm 1$. Substituting them into Eq.(\ref{solution1}), we get $W=510995.51\;eV/c^2$. And then, by Eq.(\ref{notation}), we further have $k_C,\;W_C,\;\lambda$ and $s$. Inserting all of them into Eq.(\ref{ssp16}), we obtain
\begin{eqnarray}
\nonumber \Delta E(z)\equiv(\Delta E)_{(2s^{1/2}-2p^{1/2})}&=&{Q^1(z)Q^0(z)\over R^2}\times 358.826\;eV\\
\label{ssp19} &=&{Q^1(z)Q^0(z)\over R^2}\times 8.36\times10^7\;(Lamb\;shift),
\end{eqnarray}
where $Q^0(z)\equiv ct(z)$ and $Q^1(z)$ have been given in Eqs.(\ref{newLa1}) and (\ref{new8}) respectively (see also FIG.2 and FIG.3). In the expression of function $\Delta E(z)$ of Eq.(\ref{ssp19}),  when $z$ were fixed, the only unknown number is $R$ which is the universal parameter of dS-SR. Therefore, it is expected that $R$ could be determined through accurate enough observations of the level spectrum shifts of atoms on distant galaxy. The curves of $\Delta E(z)$ of Eq.(\ref{ssp19}) with $|R|=\{0.5\times 10^5Gly,\; 10^5Gly,\;2\times 10^5Gly \}$ are shown in FIG.6. It is essential that FIG.6 shows that when $z\geq 1$, $\Delta E(z)>>1$ Lamb shift. This indicates that comparing with the usual QED's hyperfine structure effects (i.e., the Lamb shift measured in the laboratory), the dS-SR QM fine structure effects are dominating for the splitting between $2s^{1/2}$- and $2p^{1/2}$- states of Hydrogen atom on distant galaxy. In the TABLE I, the $\Delta E(z)$ for $|R|=\{10^3Gly,\;10^4Gly,\;10^5Gly\}$ and $z=\{1,\;2\}$ is listed. It is learned that $\Delta E(z)>>1$ Lamb shift to all cases too.

\begin{figure}[ht]
\begin{center}
\includegraphics[scale=1.0]{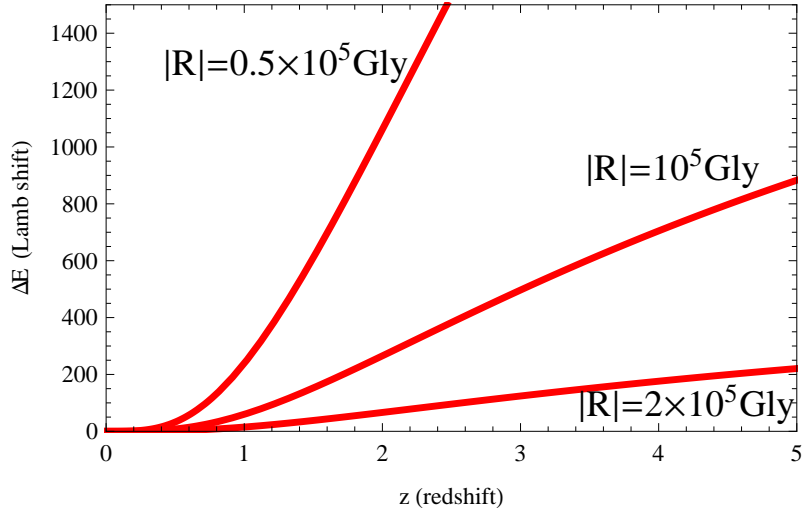}
\caption{\label{fig3}Functions $\Delta E(z)$ of
Eq.(\ref{ssp16}) with $|R|=0.5\times 10^5Gly,\; 10^5Gly,\;2\times 10^5Gly$ are shown. The unit of the energy splitting $\Delta E(z)$ is (Lamb shift)$\simeq4.3\times 10^{-6}eV$.}
\end{center}
\end{figure}

\begin{table}
\caption{ The energy level splitting of Hydrogen's $(2s^{1/2}-2p^{1/2})$, $\Delta E(z)\equiv(\Delta E)_{(2s^{1/2}-2p^{1/2})}$ (see Eqs. (\ref{ssp16}) (\ref{ssp19})): $R$ is the radius of de Sitter pseudo-sphere in dS-SR, which is a universal parameter in the theory and $|R|>R_H\equiv 13.7Gly$(Horizon of the Universe). $Gly\equiv 10^9\; light \;years$. $z$ is the red shift. (Lamb shift)$\simeq 4.3\times10^{-6}eV$.  }
\tabcolsep 0.1in
\begin{tabular}{|c|cc|cc|cc|}\hline \hline
$|R|$&\multicolumn{2}{|c|}{$10^3Gly$}&\multicolumn{2}{|c|}{$10^4Gly$}&\multicolumn{2}{|c|}{$10^5Gly$}\\ \hline
$z$&1&2&1&2&1&2\\ \hline
$\Delta E(z)\; (eV)$&2.6&11&$2.6\times10^{-2}$&0.11&$2.6\times10^{-4}$&$~~1.1\times10^{-3}$\\
(Lamb shift)&$6\times10^{5}$&$~~2.7\times10^{6}$&5977&26541&59.77&265.41\\ \hline \hline
\end{tabular}
\end{table}

\end{enumerate}

\section{Summary and discussions}

\noindent In this paper, we have solved the de Sitter special
relativistic Dirac equation of Hydrogen in the
Earth-QSO framework reference by means of the adiabatic approach and the quasi-stationary perturbation calculations of QM. Hydrogen atoms are located on the light cone of the Universe.  FRW metric and $\Lambda$CDM cosmological  model are used to discuss this
issue. To the atom, effects of de Sitter space-time geometry
described by Beltrami metric are  taken into account. The
dS-SR Dirac equation of Hydrogen turns out to be
 a time dependent quantum Hamiltonian system. We have provided an
explicit calculation to examine whether the adiabatic approach to
deal with this time-dependent system is eligible. Since the radius
of de Sitter sphere $|R|$ is cosmologically large, it  makes the
time-evolution of the system is so slow that the adiabatic
approximation legitimately works with high accuracy. Based the dS-SR Dirac equation's solutions up to $\mathcal{O}(1/R^2)$, some remarkable effects of dS-SR atom physics are revealed:
\begin{enumerate}
\item The fundamental
physics constants variate adiabatically along with cosmologic time
in dS-SR quantum mechanics framework. As is well
known that the quantum evolution in the time-dependent quantum
mechanics has been widely accepted and studied during past several
decades. It is remarkable that the
time-variations of $\mu\;(or\;m_e),\;\hbar$ and $e$ (see Eqs. (\ref{hme1}) (\ref{hme2}) (\ref{hme3})) belong to  such quantum evolution
effects.
\item The fine-structure constant $\alpha\equiv e^2/(\hbar c)$ keeps invariant along with time up to $\mathcal{O}(1/R^2)$ \cite{Yan4}. In the expression of $\alpha$, the $e^2$'s time-variation and $\hbar$'s are    canceled each other. However, whether or not this cancelation mechanism works to the next order $\mathcal{O}(1/R^4)$ remains to be open so far.

\item $(2s^{1/2}-2p^{1/2})$-splitting due to dS-SR Dirac QM effects: Distinguishing from E-SR Dirac QM theory of Hydrogen atom, the degeneracy of $ (2s^{1/2}-2p^{1/2})$ is broken in dS-SR QM. By means of the quasi-stationary perturbation theory, the $(2s^{1/2}-2p^{1/2})$-splitting $\Delta E(z)$ has been calculated analytically, which belongs to $\mathcal{O}(1/R^2)$-physics of dS-SR QM. Numerically, we found that when $|R|\simeq \{10^3 Gly,\;10^4 Gly,\;10^5 Gly\;\}$ (note the Universal horizon $R_H\simeq 13.7Gly<<|R|$), and $z\simeq \{1,\;{\rm or}\;2\}$, we have $\Delta E(z)>> 1{\rm (Lamb\; shift)}$. This indicate that for this case the hyperfine structure effects due to QED could be ignored, and the dS-SR fine structure effects are dominant. Therefore, we suggest that this effect could be used to determine the universal constant $R$ in dS-SR, and be thought as a test of new physics beyond E-SR.

\end{enumerate}

\noindent Finally, we address again that the dS-SR is a natural extension of E-SR. What we achieved in this paper are that we revealed new effects in one-electron atom physics, which are beyond E-SR and hence can be used to recognize dS-SR.

\section*{Acknowledgments}
 The work is supported in part by National
Natural Science Foundation of China under Grant Numbers 10975128,
and and by  the Chinese Science Academy Foundation under Grant
Numbers KJCX-YW-N29.

\section*{Appendix A: Electric Coulomb Law in QSO-Light-Cone Space}

\noindent Let's derive (\ref{potential-B1}). The action for deriving
electrostatic potential of proton located at
$Q^\mu=\{Q^0=ct,\;Q^1,\;Q^2=0,\;Q^3=0\}$ within background
space-time metric $g_{\mu\nu}\equiv B_{\mu\nu}(Q)$ of
eq.(\ref{new01}) in the Gaussian system of units reads
\begin{equation}\label{action1}
S=-{1\over 16\pi c}\int d^{4}L \sqrt{-g}F_{\mu\nu}F^{\mu\nu}-{e\over
c}\int d^{4}L \sqrt{-g}j^\mu A_\mu,
\end{equation}
where $g=\det (B_{\mu\nu})$, $F_{\mu\nu}={\pa A_\nu\over \pa
L^\mu}-{\pa A_\mu\over \pa L^\nu}$ and {
$j^\mu\equiv\{j^0=c\rho_{proton}/\sqrt{B_{00}},\;\mathbf{j}\}$ is
4-current density vector of proton (see, e.g, Ref.\cite{Landau}:
{\it Chapter 4; Chapter 10, Eq.(90.3)}).  The explicit
matrix-expressions for $B_{\mu\nu}(Q)$ and $B^{\mu\nu}(Q)$ up to
$\mathcal{O}(1/R^2)$ are follows:

$$
B_{\mu\nu}(Q)=\left(
\begin{array}{lccr}
1+{2(Q^0)^2-(Q^1)^2\over R^2} & -{Q^0 Q^1\over R^2} &0 &0 \\
-{Q^1 Q^0\over R^2} & -1+{2(Q^1)^2-(Q^0)^2\over R^2} & 0 & 0 \\
0 & 0 & -1-{(Q^0)^2-(Q^1)^2\over R^2} & 0 \\
0 & 0 & 0 & -1-{(Q^0)^2-(Q^1)^2\over R^2}
\end{array} \right),
$$
$$
B^{\mu\nu}(Q)=\left(
\begin{array}{lccr}
1-{2(Q^0)^2-(Q^1)^2\over R^2} & {Q^0Q^1\over R^2} &0 &0 \\
{Q^1Q^0\over R^2} & -1-{2(Q^1)^2-(Q^0)^2\over R^2} & 0 & 0 \\
0 & 0 & -1+{(Q^0)^2-(Q^1)^2\over R^2} & 0 \\
0 & 0 & 0 & -1+{(Q^0)^2-(Q^1)^2\over R^2}
\end{array} \right).
$$ }
Making space-time variable change of $L^\mu\rightarrow
L^\mu-Q^\mu\equiv x^\mu=\{x^0=ct_L-ct,\;x^i=L^i-Q^i\}$, we have
action $S$ as
\begin{eqnarray}
\nonumber S=-{1\over 16\pi c}&& \hskip-0.1in \int d^{4}x
\sqrt{-\det(B_{\mu\nu}(Q))}
F_{\mu\nu}F^{\mu\nu}-{e\over c}\int d^{4}x \sqrt{-\det(B_{\mu\nu}(Q))} j^\mu A_\mu \\
\nonumber = \left(-{1\over 16\pi c}\right.&& \hskip-0.1in
B_{\mu\lambda}(Q)B_{\nu\rho}(Q) \int  d^{4}x
F^{\mu\nu}(x)F^{\lambda\rho}(x) \\
\label{action2} && \left. -{e\over c}\int d^{4}x j^\mu (x) A_\mu (x)
\right)\sqrt{-\det(B_{\mu\nu}(Q))},
\end{eqnarray}
and the equation of motion $ \delta S/\delta A_\mu(x)=0$ as follows
(see, e.g, \cite{Landau}, Eq. (90.6), {\it pp257})
\begin{equation}\label{action3}
\pa_\nu F^{\mu\nu}=B^{\nu\lambda}\pa_\nu
F^{\mu}_{~~\lambda}=-{4\pi\over c}j^\mu.
\end{equation}
In Beltrami space, $A^\mu=\{\phi_B,\;\mathbf{A}\}$ (see, e.g.,
\cite{Landau}, eq.(16.2) in {\it pp. 45}) and 4-charge current
$j^\mu=\{c\rho_{proton}/\sqrt{B_{00}}, \;\mathbf{j}\}$. According to
the expression of charge density in curved space in Ref.
\cite{Landau}, ({\it pp.256, Eq. (90.4)}), $\rho_{proton}\equiv
\rho_B={e\over \sqrt{\gamma}}\delta^{(3)}(\mathbf{x})$ and $
\mathbf{j}=0$, where
\begin{eqnarray}\label{L1}
&&\gamma=\det (\gamma_{ij}),\\
\nonumber&&dl^2=\gamma_{ij}dx^idx^j=\left(-g_{ij}+{g_{0i}g_{j0}\over
g_{00}}\right)dx^idx^j~~~~(see~eq.(84.7)~in~Ref.[37])\\
\label{L2}&&\left.\hskip0.25in=\left(-B_{ij}+{B_{0i}B_{j0}\over
B_{00}}\right)dx^idx^j\right.
\end{eqnarray}
Noting $B_{01}=B_{10}=-{C^2t^2\over R^2}$, $B_{00}\sim 1$, and
$B_{01}B_{10}\simeq \mathcal{O}(1/R^4)\sim 0$,  we have
\begin{equation}
\sqrt{\gamma}\equiv\sqrt{\det (\gamma_{ij})}\simeq\sqrt{-\det
(B_{ij})},
\end{equation}
and hence
\begin{equation}\label{rho}
\rho_{proton}\equiv \rho_B={e\delta^{(3)}(\mathbf{x})\over
\sqrt{-\det (B_{ij})}},~~\mathbf{j}=0.
\end{equation}

\begin{enumerate}

\item  When $\mu=0$ in Eq.(\ref{action3}), we have the Coulomb's law
(\ref{potential-B}), i.e.,
\begin{eqnarray}
\nonumber
-B^{ij}(Q)\pa_i\pa_j\phi_B(x)\hskip-0.07in&=&\hskip-0.06in\left[\left(1-{(Q^0)^2-(Q^1)^2\over
R^2}\right)\nabla^2+{(Q^1)^2\over R^2}{\pa^2\over \pa
(x^1)^2}\right]\hskip-0.06in
\phi_B(x)= -{4\pi\over c} j^0\\
\label{A-1} &=&{-4\pi e\over
\sqrt{-\det(B_{ij}(Q))B_{00}(Q)}}\delta^{(3)}(\mathbf{x}),
\end{eqnarray}
where $ B^{ij}(Q)=\eta^{ij}+{(Q^0)^2-2(Q^1)^2\over R^2}
\delta_{i1}\delta_{j1}+{(Q^0)^2-(Q^1)^2\over R^2}
\delta_{i2}\delta_{j2}+{(Q^0)^2-(Q^1)^2\over R^2}
\delta_{i3}\delta_{j3}+\mathcal{O}(R^{-4})$ has been used, and
$B_{ij}$ were given in (\ref{new01}). Expanding (\ref{A-1}), we have
\begin{eqnarray*}
\left[{\pa^2\over \pa (x^1/[1+{(Q^1)^2\over 2R^2}])^2}\right. &
+&{\pa^2\over
\pa (x^2)^2} + \left.{\pa^2\over \pa (x^3)^2}\right]\phi_B(x)=-4\pi \left(1-{3(Q^0)^2-4(Q^1)^2\over 2R^2}\right) \\
 \times && \hskip-0.3in \left(1-{2(Q^0)^2-(Q^1)^2\over 2R^2}\right)\left(1+{(Q^0)^2-(Q^1)^2\over R^2}\right)
e\delta(x^1) \delta(x^2)\delta(x^3)\\
= && \hskip-0.3in -4\pi e\left(1-{3[(Q^0)^2-(Q^1)^2]\over
2R^2}\right)\delta(x^1) \delta(x^2)\delta(x^3).
\end{eqnarray*}
Noting $\delta(x^1)=\delta(x^1/[1+(Q^1)^2/2R^2])(1-(Q^1)^2/2R^2)$,
we rewrite above equation as follows
\begin{eqnarray*}
\left[{\pa^2\over \pa (x^1/[1+{(Q^1)^2\over 2R^2}])^2}\right. &
+&{\pa^2\over \pa (x^2)^2} + \left.{\pa^2\over \pa
(x^3)^2}\right]\phi_B(x)=-4\pi e\left(1-{3[(Q^0)^2-(Q^1)^2]\over
2R^2}\right) \\
 \times && \hskip-0.3in \left(1-{(Q^1)^2\over 2R^2}\right)\delta(x^1/[1+{(Q^1)^2\over 2R^2}]) \delta(x^2)\delta(x^3)\\
= && \hskip-0.3in -4\pi e\left(1-{3(Q^0)^2-2(Q^1)^2\over
2R^2}\right)\delta(x^1/[1+{(Q^1)^2\over 2R^2}])
\delta(x^2)\delta(x^3).
\end{eqnarray*}
Setting
\begin{equation}\label{AA1}
\td{x}^1\equiv x^1/[1+{(Q^1)^2\over 2R^2}],
\end{equation}
 the above equation
becomes
\begin{equation}\label{A-01}\left[{\pa^2\over \pa
(\td{x}^1)^2} +{\pa^2\over \pa (x^2)^2} +{\pa^2\over \pa
(x^3)^2}\right]\phi_B(x) =-4\pi e \left(1-{3(Q^0)^2-2(Q^1)^2\over
2R^2}\right)\delta(\td{x}^1) \delta(x^2)\delta(x^3).
\end{equation}
 Then the solution is
$\phi_B(x)=\left(1-{3(Q^0)^2-2(Q^1)^2\over 2R^2}\right)e/r_B$ with
\begin{eqnarray}\label{rB}
\nonumber r_B&=&\sqrt{(\td{x}^1)^2+(x^2)^2+(x^3)^2}\\
\nonumber &=&\left((1-{2(Q^1)^2-(Q^0)^2\over 2R^2})^2
(x^1)^2+(x^2)^2+(x^3)^2\right)^{1/2}.
\end{eqnarray}
Therefore, we have
\begin{equation}\label{A-3}
\phi_B=\left(1-{3(Q^0)^2-2(Q^1)^2\over 2R^2}\right){e\over r_B},
\end{equation}
which is the scalar potential in Eq.(\ref{potential-B1}) in the
text.

{\item When $\mu=i$ ($i=1,2,3$) in Eq.(\ref{action3}), we have
\begin{equation}\label{AB1} \pa^i\pa_\mu
A^\mu-B^{\mu\nu}\pa_\mu\pa_\nu A^i=-{4\pi\over c}j^i= 0.
\end{equation}
By means of the gauge condition
\begin{equation}\label{AB2}
\pa_\mu A^\mu=0,
\end{equation}
we have
\begin{equation}\label{AB3} B^{\mu\nu}\pa_\mu\pa_\nu A^i=0.
\end{equation}
Then
\begin{equation}\label{AB4}  A^i=0
\end{equation}
is a solution that satisfies the gauge condition (\ref{AB2}) (noting
$\pa_0A^0={\pa\over \pa x^0}\phi_B(r_B)=0$ due to ${\pa Q^0\over \pa
x^0}={\pa Q^0\over \pa L^0}=0$ ). Eq.(\ref{AB4}) is the vector
potential in Eq.(\ref{potential-B1}) in the text.}
\end{enumerate}

\section*{Appendix B: Adiabatic approximative wave functions in
$\mathcal{SR}_{cR}$-Dirac equation of hydrogen}

\noindent Now we derive the wave function of (\ref{wave1}) in the
text. We start with eq.(\ref{Dirac8}), i.e.
\begin{equation}\label{B1}
 i\hbar\pa_t\psi=H(t)\psi=[H_0(r_B, \hbar, \mu, {e})+H'_0(t)]\psi,
\end{equation}
where
\begin{eqnarray}
\label{B021} H(t)&=&H_0(r_B, \hbar, \mu, {e})+H'_0(t),\\
 \label{B2} H_0(r_B,\hbar,\mu, {e})&=& -i\hbar c
\vec{\alpha}\cdot\nabla_B + \mu c^2\beta
-{{e}^2\over r_B}~ ~(see\; eq.(\ref{Dirac4}))\\
\label{B3} H'_0(t)&=&-\left({ c^2t^2\over 2R^2}\right)
H_0(r_B, \hbar, 3\mu, e).
\end{eqnarray}

Suppose the modification of $H(t)$ along with the time change is
sufficiently slow, the system could be quasi-stationary in any
instant $\theta$. Then, in the Shr\"{o}dinger picture, the
quasi-stationary equation of $H(\theta)$
\begin{equation}\label{B5}
H(\theta)U_n(\mathbf{x}, \theta)=W_n(\theta) U_n(\mathbf{x}, \theta)
\end{equation}
can be solved. By (\ref{B021}) (\ref{B2}) (\ref{B3}) and
$t\rightarrow \theta$, the solutions are as follows (similar to
eq.(\ref{solution1}) in text)
\begin{eqnarray}\label{B06}
W_n(\theta)\equiv W_{n,\kappa}(\theta)&=&\mu_\theta c^2\left(1+{\alpha^2 \over
(n-|\kappa |+s)^2} \right)^{-1/2} \\ \nonumber
&&\alpha\equiv {e_\theta^2\over \hbar_\theta c}\simeq {e^2\over \hbar c}+\mathcal{O}({c^4\theta^4\over R^4}),~~~|\kappa|=(j+1/2)=1,\;2,\;3\;\cdots \\
\nonumber &&s=\sqrt{\kappa^2-\alpha^2},~~~~n=1,\;2,\;3\;\cdots.
\end{eqnarray}
where (see (\ref{hme1}), (\ref{hme2}), (\ref{hme3}) in text)
\begin{eqnarray}
\label{B07} &&\hbar_\theta= \left(1-{c^2\theta^2\over 2
R^2}\right)\hbar, \\
 \label{B08} &&\mu_\theta= \left(1-{(Q^1(\theta))^2\over
2R^2}\right)\mu, \\
\label{B071} && e_\theta = \left(1-{c^2\theta^2\over 4
R^2}\right)e,
\end{eqnarray}

 The complete set of
commutative observable is $\{H,\; \kappa,\;\mathbf{j}^2,\;j_z\}$, so that
we have
\begin{equation}\label{B09}
U_n(\mathbf{x},\theta)=\psi_{n, \kappa, j, j_z}(\mathbf{r}_B,\hbar_\theta, \mu_\theta, e_\theta),
\end{equation}
where $\mathbf{j}=\mathbf{L}+{\hbar\over 2} \mathbf{\Sigma},\;\hbar
\kappa =\beta(\mathbf{\Sigma}\cdot \mathbf{L}+\hbar)$. $[U_n(\mathbf{x},
\theta)]$ is complete set and satisfies
\begin{equation}\label{B10}
\int d^3x U_n(\mathbf{x}, \theta)U_m^*(\mathbf{x},
\theta)=\delta_{mn},~~~~n=\{n_r, K, j, j_z\}.
\end{equation}
 Thus, the solution of time-dependent
Shr\"{o}dinger equation (or Dirac equation) (\ref{B1}) can expanded
as follows
\begin{equation}\label{B6}
\psi(\mathbf{x}, t)=\sum_n C_n(t) U_n(\mathbf{x}, t)
\exp\left[-i\int_0^t\omega_n(\theta)d\theta\right],~~~
\omega_n(\theta)={W_n(\theta)\over \hbar}.
\end{equation}
Substituting (\ref{B6}) into (\ref{B1}), we have
\begin{equation}\label{B7}
i\hbar\sum_n(\dot{C}_nU_n+C_n\dot{U}_n)\exp\left[-i\int_0^t\omega_n(\theta)d\theta
\right]=0.
\end{equation}
By multiplying
$U^*_m\exp\left[i\int_0^t\omega_m(\theta)d\theta\right]$ to both
sides of eq.(\ref{B7}), and doing integral to $\bf{x}$ by using
(\ref{B10}), we have
\begin{eqnarray}\label{B13}
\dot{C}_m+C_m\int d^3x U_m^*\dot{U}_m
&+&\sum_n\hskip0.01in'\;C_n\int d^3xU_m^*\dot{U}_n\exp
\left[-i\int_0^t
(\omega_n-\omega_m)d\theta\right]=0, \\
\nonumber &m&=1,2,3,\cdots
\end{eqnarray}
where $\sum'_n$ means that $n\neq m$ in the summation over $n$.
Noting (\ref{B10}), we have
\begin{equation}\label{B14}
\int \dot{U}^*_m U_m d^3x+\int U_m^*\dot{U}_md^3x=0,
\end{equation}
and hence
\begin{equation}\label{B15}
\int U_m^*\dot{U}_md^3x=i\beta
\end{equation}
is purely imaginary number. Denoting
\begin{equation}\label{B16}
\alpha_{mn}=\int U_m^*\dot{U}_n d^3x,~~~~{\rm
and}~~~~\omega_{nm}=\omega_n-\omega_m,
\end{equation}
then eq.(\ref{B13}) becomes
\begin{eqnarray}\label{B17}
\dot{C}_m+i\beta C_m +\sum_n\hskip0.01in'\;C_n\alpha_{mn}\exp
\left[-i\int_0^t \omega_{nm}d\theta\right]=0.~~~~ &m&=1,2,3,\cdots
\end{eqnarray}
To further simplify it, we set
\begin{equation}\label{B18}
V_n(\mathbf{x},t)=U_n(\mathbf{x},t)\exp\left[-i\int_0^t\beta_n(\theta)d\theta\right],
\end{equation}
then
\begin{equation}\label{B19}
\psi(\mathbf{x}, t)=\sum_n C_n^{\;'}(t) V_n(\mathbf{x}, t)
\exp\left[-i\int_0^t\omega_n(\theta)d\theta\right],
\end{equation}
where
$C_n^{\;'}(t)=C_n(t)\exp\left[i\int_0^t\beta_n(\theta)d\theta\right]$,
and
\begin{equation}\label{B20}
\dot{C}_m^{\;'}(t)=[\dot{C}_m+i\beta_mC_m(t)]\exp\left(i\int_0^t\beta_n(\theta)d\theta\right)
\end{equation}
Substituting (\ref{B20}) into (\ref{B17}), we finally get
\begin{eqnarray}\label{B21}
\dot{C}_m^{\;'}+\sum_n\hskip0.01in'\;C_n^{\;'}\alpha_{mn}\exp
\left[-i\int_0^t \omega'_{nm}d\theta\right]=0.~~~~ &m&=1,2,3,\cdots
\end{eqnarray}
where
\begin{equation}\label{B22}
\omega'_{mn}=\omega'_n-\omega'_m,~~~~\omega'_n={1\over
\hbar}W_n+\beta_n.
\end{equation}
Now let's solve (\ref{B21}). Firstly, we derive $\alpha_{mn}$. By
(\ref{B5}), we have
\begin{equation}\label{B23}
{\pa H\over \pa t}U_n +H\dot{U}_n=\dot{W}_nU_n+W_n\dot{U}_n.
\end{equation}
By multiplying $U^*_m $ and doing integral over $\bf{x}$, we have
\begin{eqnarray}\nonumber
\int U^*_m\dot{H}U_nd^3x&+&\int U^*_mH\dot{U}_nd^3x=W_n\int
U^*_m\dot{U}_nd^3x \\
\label{24}{\rm
i.e.,}~~~~~\dot{H}_{mn}+W_m\alpha_{mn}&=&W_n\alpha_{mn},
\end{eqnarray}
so that
\begin{equation}\label{B25}
\alpha_{mn}=\int U^*_m\dot{U}_nd^3x={1\over
W_n-W_m}\dot{H}_{mn},~~~~ m\neq n.
\end{equation}
Therefore eq.(\ref{B21}) becomes
\begin{eqnarray}\label{B26}
\dot{C}_m^{\;'}+\sum_n\hskip0.01in'\;C_n^{\;'}{\dot{H}_{mn}\exp
\left(-i\int_0^t \omega'_{nm}d\theta\right)\over \hbar
\omega_{nm}}=0.~~~~ m=1,2,3,\cdots
\end{eqnarray}
Suppose in the initial time the system is in $s$-state, i.e.,
$C_n(0)=C_n^{\;'}(0)=\delta_{ns}$. To adiabatic process,
$\dot{H}(t)\rightarrow 0$, then the 0-order approximative solution
of eq.(\ref{B26}) is
\begin{equation}\label{B27}
[C_m^{\;'}(t)]_0=\delta_{ms}.
\end{equation}
Substituting (\ref{B27}) into (\ref{B26}), we get the first order
correction to the approximation
\begin{eqnarray}\label{B28}
[\dot{C}_m^{\;'}]_1={-\dot{H}_{ms} \over \hbar \omega_{ms}}
\exp\left(-i\int_0^t \omega'_{ms}d\theta\right)=0,~~~~ m\neq s.
\end{eqnarray}
Since the dependent on time $t$ of $U_n(t)$ is weak for adiabatic
process, eq.(\ref{B15}) indicates $\beta_n$ is small, and by
(\ref{B22}), we have $\omega'_{ms}\approx \omega_{ms}$. Then, from
(\ref{B28}), the first order correction to the solution is
\begin{eqnarray}\label{B29}
[C_m^{\;'}]_1={\dot{H}_{ms} \over i\hbar \omega_{ms}}
\left(e^{i\omega_{ms}t}-1\right),~~~~ m\neq s.
\end{eqnarray}
Substituting (\ref{B28}) (\ref{B29}) into (\ref{B19}) and neglecting
$\beta_n$, we get the wave function as follows
\begin{equation}\label{B30}
\psi(\mathbf{x}, t)\simeq  U_s(\mathbf{x}, t)e^{-i{W_s t\over
\hbar}}+\sum_{m\neq s}{\dot{H}_{ms} \over i\hbar \omega_{ms}}
\left(e^{i\omega_{ms}t}-1\right)U_m(\mathbf{x},
t)e^{\left(-i\int_0^t{W_m(\theta)\over \hbar}d\theta\right)}.
\end{equation}
By using eqs.(\ref{B09}), (\ref{B08}), (\ref{B07}), (\ref{B071}), we
finally obtain the desired results
\begin{equation}\label{wave11}
\psi(t)\simeq\psi_s(\mathbf{r}_B,\hbar_t, \mu_t,e_t)e^{-i{W_s\over \hbar}t}+\sum_{m\neq
s}{\dot{H}'(t)_{ms} \over i\hbar
\omega_{ms}^2}\left(e^{i\omega_{ms}t}-1\right)\psi_m(\mathbf{r}_B,\hbar_t, \mu_t,e_t)e^{\left(-i\int_0^t{W_m(\theta)\over
\hbar}d\theta\right)},
\end{equation}
where
\begin{eqnarray}
\label{alpha1} &&\hbar_t= \left(1-{c^2t^2\over 2
R^2}\right)\hbar, \\
\label{mu1} &&\mu_t= \left(1-{(Q^1(t))^2\over
2R^2}\right)\mu, \\
\label{mu11}&& e_t = \left(1-{c^2t^2\over 4
R^2}\right)e,
\end{eqnarray}
(\ref{alpha1}) (\ref{mu1}) and (\ref{mu11}) are  the equations
(\ref{hme1}), (\ref{hme2}) and (\ref{hme3}) in the text.  Eq.(\ref{wave11})
is just Eq.(\ref{wave1}) in the text.

\section*{ Appendix C: Calculations of elements of the perturbation Hamiltonian $\mathbf{H'}$-matric in $\mathbf{(2s^{1/2}}$-$\mathbf{2p^{1/2})}$-Hilbert space}

\noindent Now we derive Eq.(\ref{ssp3}) and Eq.(\ref{ssp9}). We start with the dS-SR Dirac spectrum equation of Hydrogen, which has been shown in Eqs. (\ref{splitting1})-(\ref{splitting4}) in the text:
\begin{equation}\label{C1}
H_{(dS-SR)}\psi=(\;H_0(r, \hbar_t, \mu_t, e_t)+H'\;)\psi=E\psi,
\end{equation}
where
\begin{eqnarray}\label{C2}
&&H_0(r,\hbar_t, \mu_t, e_t)= -i\hbar_t c
\vec{\alpha}\cdot\nabla +\mu_t c^2\beta-{e_t^2\over r},\\
\label{C3} &&H'={1\over 2}(H'^\dag+H')\equiv H'_1+H'_2,
\end{eqnarray}
where
\begin{eqnarray}
\label{C4}&&H'_1 =-{Q^1Q^0\over 4R^2}\left(\alpha^1 H_0(r,\hbar, \mu, e)+ H_0(r,\hbar, \mu, e)\alpha^1\right),\\
\label{C5}&&H'_2 ={Q^1Q^0\over 4R^2}\left(i\hbar c\overrightarrow{{\pa\over \pa{x}^1}}-i\hbar c\overleftarrow{{\pa\over \pa{x}^1}}\right).
\end{eqnarray}
The definition of $H'$-elements in the $H_0$-eigenstate space, $\langle H'\rangle_{2L^{1/2}, 2L'^{1/2}}^{m_j,\; m'_{j}}$, has been given in Eqs.(\ref{ssp1}) (\ref{ssp1-1}). The eigen values and eigen states of $H_0$ are given in the section IV.

\noindent {\bf (I) $H_1'$-matrix elements:}

\begin{enumerate}
\item  $\langle H'_1\rangle_{2s^{1/2}, 2p^{1/2}}^{1/2,\; -1/2}$:
\begin{eqnarray}\nonumber
&&\hskip-0.8in \langle H'_1\rangle_{2s^{1/2}, 2p^{1/2}}^{1/2,\; -1/2}=\int dr r^2\int d\Omega\; \psi^{1/2 \dag}_{(2s)j=1/2}(\mathbf{r})\;H'_1 \;\psi^{-1/2}_{(2p)j=1/2}(\mathbf{r})=-{Q^1Q^0\over 4R^2}W\langle (2s^{1/2})^{1/2}| \alpha^1|(2p^{1/2})^{-1/2}\rangle\\
\nonumber &&\hskip-1in =-{Q^1Q^0\over 4R^2}W\hskip-0.1in\int\hskip-0.05in dr r^2 \hskip-0.1in\int\hskip-0.05in d\Omega\hskip-0.06in \left(g_{(2s^{1/2})}(r)\chi^{1/2\dag}_{\kappa}(\mathbf{\hat{r}})_{(2s^{1/2})},
-if_{(2s^{1/2})}(r)\chi^{1/2\dag}_{-\kappa}(\mathbf{\hat{r}})_{(2s^{1/2})}\right)\hskip-0.08in
\left(\begin{array}{cc}
0 & \sigma^1\\
\sigma^1 & 0\\
\end{array} \right)\\
\nonumber &&\hskip4.0in \times\left(\hskip-0.06in\begin{array}{c}
g_{(2p^{1/2})}(r)\chi^{-1/2}_{\kappa}(\mathbf{\hat{r}})_{(2p^{1/2})}\\
if_{(2p^{1/2})}(r)\chi^{-1/2}_{-\kappa}(\mathbf{\hat{r}})_{(2p^{1/2})}
\end{array}\hskip-0.06in \right)\\
\nonumber &&\hskip-1in=-i{Q^1Q^0\over 4R^2}W\hskip-0.1in\int_0^\infty\hskip-0.1in dr r^2\left\{g_{(2s^{1/2})}(r)f_{(2p^{1/2})}(r)\hskip-0.1in\int \hskip-0.06in d\Omega\chi^{1/2\dag}_{\kappa}(\mathbf{\hat{r}})_{(2s^{1/2})}\sigma^1 \chi^{-1/2}_{-\kappa}(\mathbf{\hat{r}})_{(2p^{1/2})}\right.\\
\label{C6}&&\hskip1in \left.-f_{(2s^{1/2})}(r)g_{(2p^{1/2})}(r) \hskip-0.1in\int \hskip-0.06in d\Omega\chi^{1/2\dag}_{-\kappa}(\mathbf{\hat{r}})_{(2s^{1/2})}\sigma^1 \chi^{-1/2}_{\kappa}(\mathbf{\hat{r}})_{(2p^{1/2})}\right\}
\end{eqnarray}
where $W=W_{(n=2,\kappa=\pm 1)}$, and the explicit expressions of $2s^{1/2}$- and $2p^{1/2}$ -wave functions of
$\{g_{(2s^{1/2})}(r),\; f_{(2s^{1/2})}(r),\;g_{(2p^{1/2})}(r),\;f_{(2p^{1/2})}(r),\;\chi^{\pm 1/2}_{\pm\kappa}(\mathbf{\hat{r}})_{(2s^{1/2})},\; \chi^{\pm 1/2}_{\pm\kappa}(\mathbf{\hat{r}})_{(2p^{1/2})}\} $ are given in Eqs.(\ref{w42})$-$(\ref{ww52}) in text. From them, we have
\begin{eqnarray}
\nonumber &&\hskip-0.3in\int  d\Omega\chi^{1/2\dag}_{\kappa}(\mathbf{\hat{r}})_{(2s^{1/2})}\sigma^1 \chi^{-1/2}_{-\kappa}(\mathbf{\hat{r}})_{(2p^{1/2})}\\
\nonumber && =\int_0^{2\pi}d\phi\int_0^\pi d\theta \sin\theta \left(Y_0^0, 0\right)\left(\begin{array}{cc}
                                                                  0&1\\
                                                                  1&0
                                                                 \end{array}\right)
\left(\begin{array}{c} \sqrt{2\over 3}\cos\theta Y_1^{-1}(\theta\phi)-\sqrt{1\over3}\sin\theta e^{-i\phi}Y_1^0(\theta\phi)\\
\sqrt{2\over 3}\sin\theta e^{i\phi} Y_1^{-1}(\theta\phi)-\sqrt{1\over3}\cos\theta Y_1^0(\theta\phi)
\end{array}\right)\\
\label{C7}&& ={1\over2}\int_0^\pi d\theta \sin\theta(\sin\theta\cos\theta+\cos^2\theta)={1\over3};\\
\nonumber &&\hskip-0.3in \int  d\Omega\chi^{1/2\dag}_{-\kappa}(\mathbf{\hat{r}})_{(2s^{1/2})}\sigma^1 \chi^{-1/2}_{\kappa}(\mathbf{\hat{r}})_{(2p^{1/2})}\\
\nonumber && =\int_0^{2\pi}d\phi\int_0^\pi d\theta \sin\theta \left(-\cos\theta Y_0^0, -\sin\theta e^{-i\phi}Y_0^0\right)\left(\begin{array}{cc}
                                                                  0&1\\
                                                                  1&0
                                                                 \end{array}\right)
\left(\begin{array}{c} -\sqrt{2\over 3} Y_1^{-1}(\theta\phi)\\
\sqrt{1\over3} Y_1^0(\theta\phi)
\end{array}\right)\\
\label{C8}&&=-{1\over2}\int_0^\pi d\theta \sin\theta\cos^2\theta=-{1\over3},
\end{eqnarray}
where $Y_1^0(\theta\phi)=\sqrt{3\over 4\pi}\cos\theta,\; Y_1^1(\theta\phi)=-\sqrt{3\over 8\pi}e^{i\phi}\sin\theta,\;
Y_1^{-1}(\theta\phi)=\sqrt{3\over 8\pi}e^{-i\phi}\sin\theta$ and $Y_0^{0}(\theta\phi)=\sqrt{1\over 4\pi}$ have been used. Substituting Eqs.(\ref{C7}) (\ref{C8}) into (\ref{C6}), we get
\begin{eqnarray}\label{C9}
 \langle H'_1\rangle_{2s^{1/2}, 2p^{1/2}}^{1/2,\; -1/2}
=-i{Q^1Q^0\over 4R^2}{W\over3}\int_0^\infty dr r^2\left\{g_{(2s^{1/2})}(r)f_{(2p^{1/2})}(r)
+f_{(2s^{1/2})}(r)g_{(2p^{1/2})}(r)\right\}.
\end{eqnarray}
Inserting the explicit expressions of radial wave functions $g_{(2s^{1/2})}(r),\;f_{(2s^{1/2})}(r),$ and $
g_{(2p^{1/2})}(r),\;f_{(2p^{1/2})}(r)$ (i.e. Eqs.(\ref{43}), (\ref{w43}), (\ref{46}), (\ref{w47}) in text)
into the integral in Eq.(\ref{C9}), and accomplishing the calculations,
we have
\begin{eqnarray}\nonumber
&& \int_0^\infty dr r^2\left\{g_{(2s^{1/2})}(r)f_{(2p^{1/2})}(r)
+ f_{(2s^{1/2})}(r)g_{(2p^{1/2})}(r)\right\}\\
\label{C10}&&=\sqrt{k_C^2-W_C^2\over 4W_C^2-k_C^2}\left({W_C\over k_C}-{k_C\over 2W_C}(s+1)\right),
\end{eqnarray}
where formula $\int_0^\infty dr r^{\nu-1}\exp (-\mu r)=\Gamma(\nu)/\mu^\nu$ were used. Consequently, substituting Eq.(\ref{C10}) into Eq.(\ref{C9}), we obtain
\begin{eqnarray}\nonumber
&& \langle H'_1\rangle_{2s^{1/2}, 2p^{1/2}}^{1/2,\; -1/2}\equiv -i\Theta_1\\
\label{C11}&&=-i{Q^1Q^0\over 4R^2}{W\over3}\sqrt{k_C^2-W_C^2\over 4W_C^2-k_C^2}\left({W_C\over k_C}-{k_C\over 2W_C}(s+1)\right),
\end{eqnarray}
which is just desired result of Eq.(\ref{ssp3}), and all above calculations have been checked by the {\it Mathematica}.

\item By means of similar calculations we get also that
\begin{eqnarray}\label{C12}
\langle H'_1\rangle_{2s^{1/2}, 2p^{1/2}}^{-1/2,\; 1/2}=-i\Theta_1.
\end{eqnarray}
Since $H'_1=H'^{\dag}_1$, we have
\begin{eqnarray}\label{C13}
&&\langle H'_1\rangle_{2p^{1/2}, 2s^{1/2}}^{-1/2,\; 1/2}=
(\langle H'_1\rangle_{2s^{1/2}, 2p^{1/2}}^{1/2,\; -1/2})^*=i\Theta_1,\\
\label{C14}
&&\langle H'_1\rangle_{2p^{1/2}, 2s^{1/2}}^{1/2,\; -1/2}=
(\langle H'_1\rangle_{2s^{1/2}, 2p^{1/2}}^{-1/2,\; 1/2})^*=i\Theta_1.
\end{eqnarray}
Furthermore, to all other elements of $H'_1$-matrix, since $\int_0^{2\pi} d\phi ~\exp(\pm i n\phi)=0$ and
$\int_0^\pi d\theta \sin\theta\cos^{2n+1}\theta=0$ and etc, the explicit calculations show that all those
$H'_1$-matrix elements vanish. Consequently, all elements of $H'_1$ are calculated, and Eq.(\ref{ssp3}) is proved.
\end{enumerate}

\noindent {\bf (II) $H_2'$-matrix elements:}

 $H'_2$ has been given in Eqs. (\ref{ssp7}) (\ref{ssp8}) in the text, which is as follows
\begin{eqnarray}\label{C15}
&&H'_2 ={Q^1Q^0\over 4R^2}\left(i\hbar c\overrightarrow{{\pa\over \pa{x}^1}}-i\hbar c\overleftarrow{{\pa\over \pa{x}^1}}\right),\\
\label{C16}&&{\rm where}~~ {\pa\over\pa x^1}\equiv \pa_1=
\sin \theta\cos\phi{\pa\over\pa r}+
\cos\theta\cos\phi{1\over r}{\pa\over\pa\theta}-{\sin\phi\over r\sin\theta}{\pa\over\pa\phi}.
\end{eqnarray}

We derive Eq.(\ref{ssp9}) in text now.
\begin{enumerate}
\item $\langle H'_2\rangle_{2s^{1/2}, 2p^{1/2}}^{1/2,\; -1/2}$:
\begin{eqnarray}\nonumber
&&\hskip-0.8in \langle H'_2\rangle_{2s^{1/2}, 2p^{1/2}}^{1/2,\; -1/2}=\int dr r^2\int d\Omega\; \psi^{1/2 \dag}_{(2s)j=1/2}(\mathbf{r})\;H'_2 \;\psi^{-1/2}_{(2p)j=1/2}(\mathbf{r})\\
\nonumber &&\hskip-1.2in =-{Q^1Q^0\over 4R^2}i\hbar c\hskip-0.05in\int\hskip-0.05in dr r^2 \hskip-0.1in\int\hskip-0.05in d\Omega\hskip-0.06in \left(g_{(2s^{1/2})}(r)\chi^{1/2\dag}_{\kappa}(\mathbf{\hat{r}})_{(2s^{1/2})},
-if_{(2s^{1/2})}(r)\chi^{1/2\dag}_{-\kappa}(\mathbf{\hat{r}})_{(2s^{1/2})}\right)\hskip-0.08in
\left(\overrightarrow{\pa}_1-\overleftarrow{\pa}_1 \right)\\
\nonumber &&\hskip4.0in \times\left(\hskip-0.06in\begin{array}{c}
g_{(2p^{1/2})}(r)\chi^{-1/2}_{\kappa}(\mathbf{\hat{r}})_{(2p^{1/2})}\\
if_{(2p^{1/2})}(r)\chi^{-1/2}_{-\kappa}(\mathbf{\hat{r}})_{(2p^{1/2})}
\end{array}\hskip-0.06in \right)\\
\nonumber &&\hskip-1.2in =-i{Q^1Q^0\over 4R^2}\hbar c\int\hskip-0.05in dr r^2 \hskip-0.1in\int\hskip-0.05in d\Omega\left\{g_{(2s^{1/2})}(r)[\pa_1 g_{(2p^{1/2})}(r)]\chi^{1/2\dag}_{\kappa}(\mathbf{\hat{r}})_{(2s^{1/2})} \chi^{-1/2}_{-\kappa}(\mathbf{\hat{r}})_{(2p^{1/2})}\right.\\
\nonumber &&\hskip-1.2in +g_{(2s^{1/2})}(r)g_{(2p^{1/2})})(r)\chi^{1/2\dag}_{\kappa}(\mathbf{\hat{r}})_{(2s^{1/2})} [\pa_1\chi^{-1/2}_{-\kappa}(\mathbf{\hat{r}})_{(2p^{1/2})}]+f_{(2s^{1/2})}(r)[\pa_1 f_{(2p^{1/2})}(r)]\chi^{1/2\dag}_{\kappa}(\mathbf{\hat{r}})_{(2s^{1/2})} \chi^{-1/2}_{-\kappa}(\mathbf{\hat{r}})_{(2p^{1/2})}\\
\nonumber &&\hskip-1.2in +f_{(2s^{1/2})}(r)f_{(2p^{1/2})})(r)\chi^{1/2\dag}_{\kappa}(\mathbf{\hat{r}})_{(2s^{1/2})} [\pa_1\chi^{-1/2}_{-\kappa}(\mathbf{\hat{r}})_{(2p^{1/2})}]-[\pa_1 g_{(2s^{1/2})}(r)]g_{(2p^{1/2})})(r)\chi^{1/2\dag}_{\kappa}(\mathbf{\hat{r}})_{(2s^{1/2})} \chi^{-1/2}_{-\kappa}(\mathbf{\hat{r}})_{(2p^{1/2})}\\
\nonumber &&\hskip-1.2in - g_{(2s^{1/2})}(r)g_{(2p^{1/2})})(r)\chi^{1/2\dag}_{\kappa}(\mathbf{\hat{r}})_{(2s^{1/2})} [\pa_1\chi^{-1/2}_{-\kappa}(\mathbf{\hat{r}})_{(2p^{1/2})}]-[\pa_1 f_{(2s^{1/2})}(r)] f_{(2p^{1/2})}(r)\chi^{1/2\dag}_{\kappa}(\mathbf{\hat{r}})_{(2s^{1/2})} \chi^{-1/2}_{-\kappa}(\mathbf{\hat{r}})_{(2p^{1/2})}\\
\nonumber &&\hskip-1.2in \left.-f_{(2s^{1/2})}(r) f_{(2p^{1/2})}(r)\chi^{1/2\dag}_{\kappa}(\mathbf{\hat{r}})_{(2s^{1/2})} [\pa_1\chi^{-1/2}_{-\kappa}(\mathbf{\hat{r}})_{(2p^{1/2})}]\right\}\\
\nonumber &&\hskip-1.2in =-i{Q^1Q^0\over 4R^2}\hbar c\left\{{1\over3}\int_0^\infty dr r^2\left[{\pa g_{(2s^{1/2})}(r)\over \pa r}g_{(2p^{1/2})}(r)-{\pa g_{(2p^{1/2})}(r)\over \pa r}g_{(2s^{1/2})}(r)+{\pa f_{(2s^{1/2})}(r)\over \pa r}f_{(2p^{1/2})}(r)\right.\right.\\
\label{C17}&&\left.\left.-{\pa f_{(2p^{1/2})}(r)\over \pa r}f_{(2s^{1/2})}(r)\right]-{2\over3}\int_0^\infty dr r \left[ g_{(2s^{1/2})}(r)g_{(2p^{1/2})}(r)-f_{(2s^{1/2})}(r)f_{(2p^{1/2})}(r)\right]\right\},
\end{eqnarray}
where the integrals to $d\Omega$ have been accomplished in terms of the explicit expressions of $\chi^{1/2}_{\pm\kappa}(\mathbf{\hat{r}})_{(2s^{1/2})},\;\;\chi^{-1/2}_{\pm\kappa}(\mathbf{\hat{r}})_{(2p^{1/2})}$
in Eqs.(\ref{ww46}) (\ref{ww47}) (\ref{ww51}) (\ref{ww52}) and Eq.(\ref{C16}). Substituting expressions (\ref{43}) (\ref{w43}) (\ref{46}) (\ref{w47}) into Eq. (\ref{C17}), and finishing the integrals, we get
\begin{eqnarray}\nonumber
&& \langle H'_2\rangle_{2s^{1/2}, 2p^{1/2}}^{1/2,\; -1/2}\equiv -i\Theta_2\\
\label{C18} &&=i{Q^1Q^0\over 2R^2}{\hbar c\lambda \over 6\sqrt{4W_C^2-k_C^2}}\left({k_C^2\over W_C}-2({1\over s}+1)k_C-W_C+{2\over k_C s}W^2_C\right),
\end{eqnarray}
which is just Eq.(\ref{ssp9}), and all above result have been checked by the {\it Mathematica} calculations.

\item By means of similar calculations we get also that
\begin{eqnarray}\label{C19}
\langle H'_2\rangle_{2s^{1/2}, 2p^{1/2}}^{-1/2,\; 1/2}=-i\Theta_2.
\end{eqnarray}
Since $H'_2=H'^{\dag}_2$, we have
\begin{eqnarray}\label{C20}
&&\langle H'_2\rangle_{2p^{1/2}, 2s^{1/2}}^{-1/2,\; 1/2}=
(\langle H'_2\rangle_{2s^{1/2}, 2p^{1/2}}^{1/2,\; -1/2})^*=i\Theta_2,\\
\label{C21}
&&\langle H'_2\rangle_{2p^{1/2}, 2s^{1/2}}^{1/2,\; -1/2}=
(\langle H'_2\rangle_{2s^{1/2}, 2p^{1/2}}^{-1/2,\; 1/2})^*=i\Theta_2.
\end{eqnarray}
Furthermore, to all other elements of $H'_2$-matrix, since $\int_0^{2\pi} d\phi ~\exp(\pm i n\phi)=0$ and
$\int_0^\pi d\theta \sin\theta\cos^{2n+1}\theta=0$ and etc, the explicit calculations show that all those
$H'_2$-matrix elements vanish. Consequently, all elements of $H'_2$ are calculated, and Eq.(\ref{ssp9}) is proved.
\end{enumerate}


\end{document}